\documentclass[12pt,preprint]{aastex}
\shorttitle{GIs, Chondrules, and FU Ori}
\shortauthors{Boley and Durisen}

\begin{document}
\title{Gravitational Instabilities, Chondrule Formation, and the FU Orionis Phenomenon}

\author{Aaron C.\ Boley\altaffilmark{1}\altaffilmark{2} \&
Richard H.~Durisen\altaffilmark{1}}

\altaffiltext{1}{Astronomy Department, Indiana University, 727 E.\ Third St.,
Bloomington, IN 47405-
7105}
\altaffiltext{2}{Institute for Theoretical Physics, University of Zurich, Winterthurerstrasse 190,
Zurich, CH-8057, Switzerland; acboley@physik.uzh.ch}

\begin{abstract}
Using analytic arguments and numerical simulations, we examine whether chondrule formation and the FU Orionis phenomenon can be caused by the burst-like onset of gravitational instabilities (GIs) in dead zones.  At least two scenarios for bursting dead zones can work, in principle.  If the disk is on the verge of fragmention, GI activation near $r\sim4$ to 5 AU can produce chondrule-forming shocks, at least under extreme conditions.  Mass fluxes are also high enough during the onset of GIs to suggest that the outburst is related to an FU Orionis phenomenon.  This situation is demonstrated by numerical simulations.  In contrast, as supported by analytic arguments, if the burst takes place close to $r\sim1$ AU, then even low pitch angle spiral waves can create chondrule-producing shocks and outbursts.  We also  study the stability of the massive disks in our simulations against fragmentation and find that although disk evolution is sensitive to changes in opacity, the disks we study do not fragment, even at high resolution and even for extreme assumptions.  

\end{abstract}


\section{Introduction}

\subsection{GIs, MRI, FU Orionis Outbursts, and Chondrules}

Gravitational instabilities can activate in a disk when the Toomre (1964) parameter $Q={c_s\kappa}/{\pi G\Sigma}\lesssim 1.7$ for a thick disk (see Durisen et al.~2007), where $c_s$ is the sound speed, $
\Sigma$ is the surface density, and $\kappa$ is the epicyclic frequency.   As indicated by $Q$, a disk is unstable against GIs when it is cold and/or massive.  The resulting spiral waves driven by self-gravity efficiently transfer angular momentum outward and mass inward (e.g., Lynden-Bell \& Kalnajs 1972; Durisen et al.~1986).  Another mechanism that can efficiently transfer angular momentum outward is the magnetorotational instability (MRI; see Balbus \& Hawley 1991; Desch 2004).  In contrast to GIs, the MRI only requires a weak magnetic field coupled to the gas.  These mechanisms, either separately or in combination, are likely to be the principal way T Tauri stars accrete gas from a disk (Hartmann et al.~2006).

In order for the MRI to occur, ionized species must be present in the gas phase.   Thermal ionization of alkalis occurs wherever $T\gtrsim1000$ K, but depletion of ions by dust grains may move the temperature threshold closer to $T\sim1700$ K (Desch 1999; Sano et al.~2000), where the dust sublimates completely. Elsewhere, the ionization must be driven by a nonthermal source, e.g., energetic particles (EPs).  For this discussion, EPs refers to any particles that could ionize the gas, e.g., X-rays.  Gammie (1996) proposed that disks may have active and inactive MRI layers due to attenuation of EPs by the gas.  In the inner regions of a disk  (at radii $\sim$ few AU) where the column densities are large, MRI may only be active in a thin layer, resulting in {\it layered accretion}.  As one moves outward, column densities drop, and the entire disk can become MRI active.  The region where the MRI is mostly absent is called the {\it dead zone}.   EPs are attenuated by a surface density of only about 100 g cm$^{-2}$ (Stepinski 1992), and so even a minimum mass solar nebula (MMSN) will likely exhibit layered accretion (Desch 2004).  Even if mass accretion is only reduced and not altogether halted as a result of Reynolds stresses (Fleming \& Stone 2003; Oishi et al.~2007), mass may still pile up in the dead zone.  If enough mass accumulates, then even for an otherwise low-mass disk, GIs can activate.

The FU Orionis phenomenon is
characterized by a rapid (1-10s yr) increase in optical brightness of a young T Tauri
object, typically by 5 magnitudes, and is driven by
sudden mass accretion of the order $10^{-4} M_{\odot}~\rm yr^{-1}$ from the inner disk
onto the star (Hartmann \& Kenyon 1996). 
Because FU Ori objects appear to have
decay timescales of about 100 yr, the entire mass of a MMSN ($\sim0.01~M_{\odot}$) can be accreted onto the
star.  To date, the best explanation for the
optical outburst is a thermal instability (e.g., Bell \& Lin 1994; Kley \& Lin 1999; see discussions in Hartmann \& Kenyon 1996, Green et al.~2006; and Zhu et al.~2007).
 Armitage et al.~(2001) suggested that GIs in a bursting dead zone 
might be able to trigger an
FU Ori outburst by rapidly increasing the accretion into the inner disk ($\lesssim 0.1$ AU) and initiating an MRI through thermal ionization.
Hartmann (2007, in private communication) and Zhu et al.~(2007) also suggest that the heating due to gravitational torques might drive an MRI for $r>0.1$ AU, which would then feed mass inside 0.1 AU until a thermal instability sets in.   The FU Ori phenomenon may be a result of a {\it cascade} of instabilities, starting with a burst of GI activity in a dead zone, followed by accretion due to an MRI, followed finally by a thermal instability (cf Kley \& Lin 1999).  Indeed, recent observations of FU Ori indicate that very large mass fluxes are present out to at least $r\sim0.5$ AU (Zhu et al.~2007).  

Although details are still being debated, most chondrules formed in the first 1 to 3 Myr of the Solar Nebula's evolution (Bizzarro et al.~2004; Russell et al.~2005).   Chondrule precursors were flash melted from solidus to liquidus, where high temperatures $T\sim1700$ K were experienced by the precursors for a few minutes. The melts then cooled over hours, with the actual cooling time depending on chondrule type. 
 Chondrule collisional histories and isotopic fractionation data, chondrule-matrix complementarity, fine-grained rim accumulation, and petrological and parent body location arguments  (Krot et al.~2005) suggest that chondrules formed in the Solar Nebula (Wood 1963) in strong, localized, repeatable heating events.  The shock wave model for chondrule formation can accommodate these observational constraints and reproduce heating and cooling rates required to form chondrule textures (Iida et al.~2001; Desch \& Connolly 2002; Cielsa \& Hood 2002; Miura \& Nakamoto 2006). 
One plausible source of chondrule-producing shocks is a global spiral wave (Wood 1996).  Harker \& Desch (2002) suggest that spiral waves could also explain thermal processing at distances as large as 10 AU, which may be necessary to explain  observations of comets (Wooden et al.~2005) and recent {\it Stardust} results (e.g., McKeegan et al.~2006). It has been suggested that bursts of GIs may be able to produce the required shock strengths  (Boss \& Durisen 2005) and provide a source of turbulence and mixing (Boss 2004b; Boley et al.~2005).  Global spiral shocks are appealing because they fit many of the constraints above.  They may be repeatable, depending on the formation mechanism for the spiral waves; they are global, but produce fairly local heating; they can form chondrules in the disk; and they can work in the inner disk as well as the outer disk.

\subsection{Fragmentation}

Knowing under what conditions protoplanetary disks can fragment is crucial to understanding disk evolution inasmuch as a fragmented disk may produce gravitationally bound clumps.  This has become known as the {\it disk instability}  hypothesis  for the formation of gas giant planets (Kuiper 1951; Cameron 1978; Boss 1997, 1998).   The strength of GIs is regulated by the cooling rate in disks (Tomley et al.~1991, 1994; Pickett et al.~1998, 2000, 2003), and if the cooling rate is high enough in a low-$Q$ disk, a disk can fragment (Gammie 2001). Gammie quantified that a disk will fragment when $t_{\rm cool}\Omega\lesssim 3$ for a disk with a  $\Gamma=2$, where $\Gamma$ is the two-dimensional adiabatic index, such that  $\int p dz = P\sim\Sigma^{\Gamma}$, where $p$ is the gas pressure and $z$ is the vertical direction in the disk. Here,  $t_{\rm cool}$ is the local cooling time and $\Omega$ is the angular speed of the gas.  This criterion was approximately confirmed in 3D disk simulations by Rice et al.~(2003) and Mej\'ia et al.~(2005).  Rice et al.~(2005) showed through 3D disk simulations that this fragmentation criterion depends on the 3D adiabatic index $\Gamma_1$ and, for $\Gamma_1=\gamma=5/3$ or 7/5, the fragmentation limit occurs when $t_{\rm cool}\Omega\lesssim6$ or 12, respectively.  These results show that a change by a factor of about 1.2 in $\gamma$ has a factor of two effect on the critical cooling time.  In addition, these results indicate that the cooling time must be roughly equal to the dynamical time of the gas for the disk to be unstable against fragmentation when $\gamma=5/3$. 

 Do such prodigious cooling rates occur in disks when realistic opacities are used with self-consistent radiation physics? This question is heavily debated in the literature (e.g., Nelson et al.~2000; Boss 2001, 2004a, 2005; Rafikov 2005, 2007; Boley et al.~2006, 2007b; Mayer et al.~2007; Stamatellos \& Whitworth 2008).  The simulations to date use a wide variety of numerical methods, including very different approximations for radiation physics, and only two groups have published results of radiative transfer tests appropriate for disks (Boley et al.~2007b; Stamatellos \& Whitworth 2008). 

Nelson et al.~(2000) used 2D SPH simulations with radiation physics to study protoplanetary disk evolution.  Because their simulations were evolved in 2D, they assumed that the disk at any given moment was in vertical hydrostatic equilibrium.  Using a polytropic vertical density structure and Pollack et al.~(1994) opacities, they cooled each particle according to an appropriate effective temperature.  In their simulations, the cooling rates are too low for fragmentation.   In contrast, Boss (2001, 2005) employed radiative diffusion in his 3D grid-based code; fragmentation occurs in his simulated disks.  Besides the difference in dimensionality of the simulations, Boss assumed a fixed temperature structure for Rosseland mean optical depths less than 10, as measured along the radial coordinate (cf recent flux-limited simulations by Boss 2008). Boss (2002) found that the fragmentation in his disks is insensitive to the metallicity of the gas and attributed this independence to fast cooling by convection (Boss 2004a).  However, it must be noted that Nelson et al.~(2000) assumed a vertically polytropic density structure.  Because the specific entropy $s\sim \ln K$, where $p=K\rho^{\gamma}$ is the polytropic equation of state and $\rho$ is the gas density, the Nelson et al.~approximation effectively assumes efficient convection. 

Except for extremely massive and extended disks (Stamatellos et al.~2007), recent simulations with radiative physics by Cai et al.~(2006), Boley et al.~(2006, 2007b), and Stamatellos \& Whitworth (2008) find long cooling rates and no efficient cooling by convection.  Cai et al.~also show that the strength of GIs is dependent on the metallicity.  Furthermore, Boley \& Durisen (2006) suggest that shock bores, which can cause a rapid vertical expansion in the post-shock region of spiral shocks, could be misidentified as ``convective'' flows.  In both contrast and support of these studies, Mayer et al.~(2007) use SPH simulations with 3D flux-limited diffusion and find that fragmentation only occurs once the mean molecular weight of the gas $\mu\gtrsim2.7$. However, the simulations presented by Mej\'ia (2004), Cai et al.~(2006), and Boley et al.~(2006) unintentionally were evolved with $\mu\approx2.7$ due to an error in the inclusion of He in the opacity tables, and their disks do not fragment.  The issue of fragmentation in radiatively cooled disks thus remains unsettled.

\subsection{Current Study}

For this study, we adopt the hypothesis, which we refer to as the {\it unified theory}, that bursts of GI activity in dead zones drive the FU Ori phenomenon and produce chondrule-forming shocks.  This hypothesis is an amalgamation  of ideas presented in Wood (1996, 2005), Gammie (1996), Armitage et al.~(2001), and Boss \& Durisen (2005). In order to investigate this scenario, we designed a numerical experiment to evolve a
massive, highly unstable disk with an initial radial extent between 2 and 10 AU.  Commensurately, we investigate the stability of these massive, gravitationally unstable disks against fragmentation to assess the feasibility of the disk instability hypothesis for gas giant formation.

\section{Expectations}

The shock speed $u_1$ for a fluid element entering a global, logarithmic spiral wave with pitch angle $i$ is approximately described by
\begin{equation}
u_1\approx30 {\rm~km~s^{-1}} \left(\frac{M}{M_{\odot}}\right)^{1/2}\left(\frac{r}{\rm AU}\right)^{-1/2}\bigg| 1-\left(\frac{r}{r_p}\right)^{3/2}\bigg|\sin i,
\end{equation}
where $r_p$ is the corotation radius of the spiral pattern and where we have assumed that the gas azimuthal motion is Keplerian and $v_r$ is negligible.
When $r_p\gg r$, the shock speed limits to the Keplerian speed times $\sin i$; whether spiral waves with large $r_p$ produce chondrules is mainly dependent on the pitch angle of the spiral wave.  In contrast, when $r\gg r_p$, the speed increases
as $r$, and even shallow pitch angles can produce chondrules. 
For example, consider the MMSN, where the midplane $\rho(r)=1.4\times10^{-9} \left(r/{\rm 1~AU}\right)^{-11/4}$ g cm$^{-3}$ (Hayashi et al.~1985). Figure 1 shows the $u_1$-$\rho$ plane with heavy, solid curves indicating shock speeds for
$r_p=1$, 2.5, 5,  and $\infty$ AU and with $i=10$ and 30$^{\circ}$.  The colored regions on the plot highlight where chondrule formation is expected in a MMSN (yellow) and a 10$\times$MMSM (orange).  The chondrule-forming curves are based on the results of Desch \& Connolly (2002), which we summarize by $u_1\approx-\left(11+2\log \rho \left({\rm g~cm^{-3}}\right)\right) \pm0.5{\rm~km~s^{-1}}$.  The boundaries set by these curves are meant to be illustrative, not definitive.  Spiral shocks along shallow pitch angle spiral waves ($i\approx 10^{\circ}$) are mostly if not entirely out of the chondrule-forming range between $r=1$ and 5 AU for shocks inside corotation. However, spirals with corotation in the inner disk can produce chondrule-forming shocks for a wide range of radii, depending on the actual $r_p$.   If the spiral waves have very open pitch angles ($i\gtrsim 30^{\circ}$), then spiral waves with corotation near $r=5$ AU can produce chondrule-forming shocks near the asteroid belt and at cometary distances. The mass of the disk does not change these general behaviors. A major caveat for this simple-minded approach is that the fluid elements' motions may be poorly described by equation (1) if vertical and radial excursions induced by shock bores (Boley \& Durisen 2006) cannot be neglected.  

The possibility of producing chondrules in the asteroid belt and at comet distances during the same burst is attractive.  Moreover, we would also like to use these simulations to study other phenomena such as radiation transport, convection, and disk fragmentation in the 2-10 AU region.  For these reasons, we tried to design a numerical experiment  to investigate the connection between GIs, FU Orionis events, and chondrule formation that is biased toward producing strong shocks with a corotation near 5 AU and with open pitch angles.  

\section{Methodology}

\subsection{Initial Conditions}

If chondrule-producing
shocks cannot be created in simulations biased in their favor, then it would present a serious
problem for GIs as the source of chondrule processing.   The reader should keep in mind that the model, at this stage, is not necessarily meant to be representative of a typical disk, although the massive disk we present may be plausible for early Class I objects (e.g., L1551; Osorio et al.~2003). To create the initial conditions (ICs), consider a disk that is vertically polytropic, is axisymmetric, has a constant
$\gamma$, and is Keplerian ($\kappa=\Omega$).  Also assume that self-gravity is negligible. With these assumptions, the vertical disk structure can be
calculated by using the following equations:
\begin{eqnarray}\rho(z) &=& \rho_0\left(1-z^2/h^2\right)^{1/(\gamma-1)},\\ 
h^2 & = & \frac{2\gamma K}{\Omega^2(\gamma-1)}\rho_0^{\gamma-1},{\rm~and}\\ 
\rho_0 &=&\left(\Sigma \Omega\left(\frac{\gamma-1}{2\gamma \pi K}\right)^{1/2}
\frac{\Gamma[(3\gamma-1)/(2\gamma-2)]}{\Gamma[\gamma/(\gamma-1)]}\right)^
{2/(\gamma+1)},
\end{eqnarray}
where $K$ for $p=K\rho^{\gamma}$ is the polytropic coefficient at any given $r$, $G$ is the
gravitational constant, $\Omega$ is the Keplerian angular speed, and $h$ is the disk scale height.
Using equations (2) through (4), one can show that
\begin{eqnarray}\Sigma(r)&=&\pi^{-(3\gamma+1)/4}\left(\frac{2}{\gamma-1}
\left(\frac{\Gamma[\gamma/(\gamma-1)]}{\Gamma[(3\gamma-1)/(2\gamma-1)]}\right)^2
\right)^{(1-\gamma)/4}\\\nonumber &\times&\left(\gamma
K(r)\right)^{1/2}\left(GQ(r)\right)
^{-(\gamma+1)/2}\Omega(r)^\gamma.\end{eqnarray}
  To select a 
model, one needs to specify power laws for $\Sigma$ and $Q$, $\Sigma$ and $K$, or
$K$ and $Q$.   For calculating $Q$, the midplane sound speed $c_s$ is used.
Note that when $Q=$ constant, there exists a lower limit for $q$, $\Sigma\sim r^{-q}$, where the radial entropy gradient remains positive.  Because the specific entropy $s\sim \ln K$, equation (5) can be solved to show that $ds/dr\sim (3\gamma-2q)/r$ for constant $Q$.  The critical surface density power law is then $q_c=5/2$ for $\gamma=5/3$ and $q_c=21/10$ for $\gamma=7/5$.  When $q>q_c$, radial thermal convection may set in for a constant $Q$ disk. However, the radial entropy gradient is not the only stability criterion for radial thermal convection in disks (Klahr \& Bodenheimer 2003).

Equations (2) through (5) are used to generate the
ICs for this study.  The disk is massive, $0.09~M_{\odot}$, with $\Sigma\sim r^{-2}$, initially between about 2 and 10 AU in radius and with a $Q\sim2$ everywhere.  Although the surface density power law is steep, it is consistent with the Nice-MMSN (Desch 2007).\footnote{This model for the MMSN is derived  in a way similar to Weidenschilling's (1977) approach; but, instead of using the current positions of Solar System planets, Desch uses initial positions based on the Nice model (e.g., Levison et al.~2007). }  Because the analytic model ignores self-gravity, assumes a constant $\gamma$ (see \S 3.2 below), and has sharp disk truncation edges, the model must be relaxed to a new equilibrium structure in the hydro code.
 Once the ICs are generated, they are loaded into our code and evolved without cooling at very low azimuthal resolution, essentially two-dimensionally.  The radial and vertical momenta are damped to allow the disk to relax calmly to an equilibrium configuration, which takes about ten orbits at 10 AU.   To create a mass concentration in the disk, mass is then added to
the disk linearly in time with a Gaussian profile. The specific internal energy is held constant.
  The peak of the Gaussian is centered on 5
AU.  The radial
FWHM of the Gaussian is chosen to be 3 AU, which is roughly the size of the
most unstable radial wavelength $\lambda_u\approx 2\pi c_s/Q\kappa\approx2\pi h/Q\approx 0.2\pi r$ (Binney \& Tremaine 1987). 
Mass is added until nonaxisymmetry is visible in midplane density images, which
takes approximately 190 yr, i.e., 6 orp of evolution (1 orp is defined as 1 outer
rotation period at 10 AU for this model, which is about 33 yr). The total mass added is
about 0.08 $M_{\odot}$, and so if one were to imagine a corresponding accretion
rate, it would be about $4\times 10^{-4}~M_{\odot}$ yr$^{-1}$. It is probably
unphysical to add so much mass so quickly to the disk without increasing the temperature.  However, we remind the reader that the study is 
a numerical experiment.  Figures 2 and 3 illustrate the density distributions of the disk before and after the mass accretion.

The mass buildup increases the total mass of the simulated disk to $\sim0.17$ M$_{\odot}$ and 
it decreases $Q$ below unity in a narrow region around 5 AU just
before strong GI activation. Because the ring is grown over 6 orp, 
the instability is not obviously overshot. The mass-weighted average $Q_{\rm av}$ over
the FWHM of the Gaussian centered at 5 AU is approximately unity when the ring
becomes unstable.  This implies that $Q$ must approach unity for a
spatial range of at least the most unstable radial wavelength before GIs will activate in a
disk.

\subsection{CHYMERA}

The disk is evolved with the CHYMERA code (Boley 2007), which employs the Boley et al.~(2007b) radiation physics algorithm.   CHYMERA is an Eulerian code that solves the hydrodynamics equations on a fixed, cylindrical grid.  The code includes self-gravity, a ray+flux-limited diffusion hybrid scheme, and a variable first adiabatic index $\Gamma_1$. This code has been extensively tested. In response to criticisms of Boss's (2001, 2002, 2004a, 2005) simulations by Boley et al.~(2007a), Boss (2007) questions the accuracy of the Indiana University code, which is the predecessor of CHYMERA.   Boss raises concerns that we address here.

{\it Potential Solver:} CHYMERA uses a direct Poisson solver to calculate the potential due to mass on the grid. To provide the boundary conditions for the direct solver, a spherical harmonics expansion for $l=\vert m\vert\lesssim10$ is used to calculate the potential on the grid boundary (e.g., Pickett et al.~2003; Boley et al.~2007b) .  Boss argues that the boundary solver used in CHYMERA may be too low-order to capture clump formation.  Tests by Pickett et al.~(2003) and Mej\'ia et al.~(2005) and the results from the Wengen Test 4 code comparison (WT4; Mayer et al.~2008, in prep.) demonstrate that this is not the case.  During the initial analyses of WT4, the potential solver was rigorously tested in an attempt to explain some deviations noticed between CHYMERA and other codes.  These deviations were finally attributed to differences in the initial conditions, and new simulations with consistent ICs show very good agreement between codes.  To test the potential solver in CHYMERA, a high-order Bessel function expansion boundary solver (Cohl \& Tohline 1999) was employed with Legendre half-integer polynomials expanded out to $m=30$, which showed convergence to better than $10^{-6}$ when compared with $m=40$.  The spiral structure and clump formation in the simulation with this high-$m$ boundary solver were indistinguishable from the normal lower-order spherical harmonics expansion.  The Wengen test is very demanding, with strong nonaxisymmetric structure and a highly flattened disk, and yet the lower-order spherical harmonics expansion for $l=\vert m\vert\lesssim10$ used in CHYMERA (e.g., Boley et al.~2007b) captures clump formation.  

{\it Resolution \& Numerical Heating:}  The resolution for the high-resolution simulations presented in this paper are at higher resolution than the base simulations presented for Wengen Test 4. As will be demonstrated here, CHYMERA shows fragmentation when it is expected based on experimentally determined fragmentation criteria (Gammie 2001; Rice et al.~2005).  The numerical heating reported by Boley et al.~(2006) affected a small region of their inner disk, and is understood to be a resolution effect.  The simulations presented here are at high enough resolution, as determined by the Boley et al.~(2007b) tests, that this effect should be negligible if present at all. 

{\it Radiative Transfer Boundary Conditions:} CHYMERA has been shown to relax to analytic solutions for the flux through an atmosphere as well as the temperature profile for high, moderate, and low optical depths.  The code allows convection when it should, and does not when it should not (Boley et al.~2007b).

For the treatment of H$_2$,  we use a
frozen ortho/para ratio of 3:1 in this study for reasons presented in Boley et al.~(2007a).
 The grain size
distribution is assumed to follow the ISM power law, $dn \sim a^{-3.5} da$
(D'Alessio et al.~2001), where the maximum grain size $a_{\rm max}$ is chosen to be
1 mm to account for grain growth (D'Alessio et al.~2006).  To transition smoothly between the low and high optical depth regimes, a weighted combination between Rosseland and Planck mean opacities $\kappa_w$ is used
$\kappa_w=(\kappa_{\rm R} \tau_R + \kappa_P/\tau_R)/(\tau_R+1/\tau_R)$, where $\tau_R$ is the vertical midplane Rosseland optical depth.  Different methods for interpolating $\kappa_R$ and $\kappa_P$ {\it along the ray} are being explored, but they have not been implemented here. The particles are assumed to be well-mixed with the
gas.  The effect of dust settling is heuristically considered by
evolving the disk in several ways: standard opacity and $10^{-2}$, $10^{-3}$, and $10^{-4}$ of the
standard opacity, referred to as the $10^{-2}~\kappa$, $10^{-3}~\kappa$, and $10^{-4}~\kappa$ simulations. This choice in opacities varies the vertical midplane Rosseland mean optical depths from about $\tau=10^4$ to unity, and spans the limits where advection of photons becomes important ($\tau v/c>1$;  Krumholz et al.~2007) and where radiative cooling becomes most efficient ($\tau\sim1$). No external radiation
is assumed to be shining onto the disk, except for a 3 K background temperature.
Although such rapid dust settling and a naked disk are likely to be unrealistic, the assumptions are made to bias the disk toward strong shocks
inasmuch as irradiation and opacity affects GI strength (Cai et al.~2006, 2008).  For a base comparison, an adiabatic simulation, i.e., where shock heating is allowed but not cooling, is evolved for about 33 yr.

The simulations are evolved at two resolutions:
$(r,~\phi,~z)=256\times512\times64$ and $512\times1024\times128$ cells, respectively, above the midplane.   
The lower resolution simulations (512 sim) have a grid spacing of 0.05 AU per cell in $r$
and $z$ and the higher 0.025 AU (1024 sim).  For both resolutions, $r\Delta \phi\approx \Delta r$ at about $r=4$ AU.  These parameters are summarized in Table 1.
 For each disk, 1000 fluid tracer elements are randomly
distributed in $r$ between about 3 and 7 AU, in $\phi$ over $2\pi$, and in $z$
roughly within the
scale height of the disk. 

Due to inefficient cooling in the standard simulations, high enough temperatures in
the disk are reached after 80 and 55 yr for the 512 and 1024 simulations, respectively, to create radiative timescales that are too short to resolve;  the
simulations are stopped.  As described below, the reduced
opacity simulations are able to cool much more efficiently, and the disk
temperatures remain manageable with the time-explicit algorithms used in CHYMERA.
However, the reduced opacity simulations are stopped after about 110 yr due to
the development of a strong, one-arm spiral, which is treated incorrectly with our
fixed-star assumption. 

\subsection{Fluid Element Tracer}

The hydrodynamics scheme in CHYMERA is Eulerian, and does not give direct information on the histories of tracer fluid elements.  In order to derive detailed and statistical shock information and to capture the
complex gas motions in unstable disk simulations, we have combined a tri-Akima spline interpolation algorithm
with a Runge-Kutta integrator (e.g.,
Press 1986) 
for tracing a large sample of fluid elements during the simulation.  Velocities and thermodynamic quantities,
e.g., temperature and density,
are interpolated once each time step. Because the hydrodynamics code 
explicitly solves the equation of motion for the gas, the algorithm only needs 
time and velocity information to integrate the positions of the fluid elements.  The Runge-Kutta integrator was intended to be fourth-order accurate; unfortunately, an implementation error when combining the integration scheme into CHYMERA resulted in a first-order accurate scheme in time.  Because tracing fluid element histories is an important point of this paper, we reran a stretch of one of the simulations with a second-order fluid element tracing routine.  After about one orbit at $r\sim5$ AU, the mean fractional differences, $(x_1-x_2)/x_2$, between the first and second-order schemes are  $-3\times 10^{-4}$, $-2\times 10^{-4}$,
and $-10^{-3}$ for $r$, $\phi$, and $z$, respectively.  The sample standard deviations are 
$4\times 10^{-3}$, $10^{-2}$, and $10^{-1}$ for $r$, $\phi$, and $z$, respectively.  The  mean differences are small, and show that there are no obvious systematic effects.  The sample deviations
indicate that the variations between the the two schemes are also marginal for the radial and azimuthal directions.  There is considerably larger scatter in the vertical direction, where spiral shocks
can create complicated vertical flows (Boley \& Durisen 2006).  Because the fluid
elements are used to gather shock information, the scatter in the vertical positions between the two schemes is not a 
major concern for this study.  The effect is similar to redistributing the fluid elements vertically, keeping the same $r$ and $\phi$, every few orbits.  

An Akima spline is similar to a natural cubic spline, but typically yields better results
for curves with sudden changes (see Akima 1970), as are expected in shock profiles.  
Although the spline fits a curve to a one-dimensional set of data points,
the interpolation can be extended to data in three dimensions.  First, consider a
cubic volume with data at the vertices. A value anywhere in the volume can be
approximated by seven linear interpolations: four to calculate values between each vertex in 
a particular direction, two to calculate the values along the projections of the desired point onto the
interpolated lines, and a final interpolation through the point of
interest.  Extending this
from a simple tri-linear fit to a tri-Akima spline is relatively
straightforward.  Instead of using the eight nearest points that enclose a volume, 
one uses the 125
closest points, with five data points used for every Akima spline fit. The central
data point is the closest data value to the point of interest. We use the 
GNU Scientific Library Akima spline algorithm for performing fits.  

\section{Results}

\subsection{Structure}

As these disks become gravitationally unstable, low-order spiral waves develop.  These waves drive a sudden increase in mass flux throughout the disk and drive strong spiral shocks.  Following Mej\'ia et al.~(2005), we refer to this phase of disk evolution as a burst of GI activity.
Surface density plots for the 512 simulations are shown in Figures 4 and 5 at
$t\approx 33$ and 77 yr, respectively.  
 The images at $t\approx33$ yr indicate that except for the $10^{-4}~\kappa$ simulation, the burst
has a well-defined, three-arm spiral and that the spiral arms become denser
as the opacity is lowered.  By 77 yr, the disk structures are noticeably different.
Each disk has a visually distinct two-arm spiral, again, except for the $10^{-4}$ $\kappa$ simulation. 
  For comparison with Figures 4 and 5, the 1024 simulations are shown in Figures 6 and 7 at $t\approx33$  and 78 yr, respectively.
The higher resolution simulations have additional fine structure, but they do not develop clumps or dense knots.  In fact, some of the knotty structure that develops in the 512 10$^{-4}$ $\kappa$ simulation is absent at higher resolution.  
The lower opacity
simulations appear to have stronger amplitudes, with denser spiral arms and stronger mid-order Fourier components (see below).  As will be discussed, the $10^{-4}$ $\kappa$ simulation is on the verge of fragmentation. 

The visual differences between the simulations can be quantified by a Fourier mode
analysis. Define $A_m\equiv 2 \int (a_m^2+ b_m^2)^{1/2} r dr /\int a_0 r dr $, where $\pi a_m=\int \Sigma \cos (m\phi) d\phi$ and $\pi b_m=\int \Sigma \sin(m\phi) d\phi$ (cf.~Boley et al.~2006 who use the volume density). In addition, define $A_+= \sum_{m=2}^{32} A_m$. As shown in Table 1,  $\left<A_+\right>$ tends to increase with both resolution and decreased opacity.  This is also seen in the Fourier component spectrum, shown in Figure 8, where squares indicate the 1024 simulations and crosses lines the 512 simulations.    The $10^{-3}~\kappa$ simulation is not shown for readability, but if falls between the $10^{-2}$ and $10^{-4}~\kappa$ simulations and has the least amount of divergence between the 512 and 1024 simulations at large $m$. For all runs, the Fourier $m=1$ through 6 dominate, and the power in the $A_m$ spectra drops off steeply around $m\sim10$.  The 1024 simulations have a shallower profile at large $m$, so resolution effects probably play a role for $m\gtrsim10$, even at resolutions of LMAX=512.      The convergence of the $A_m$ profile for large $m$ is a topic for future investigation (Steinman-Cameron et al., in preparation).

\subsection{Energy Budget}

The energetics of the disk simulations are shown in Figures 9 and 10, where
cumulative energy loss by radiation (blue) and cumulative energy dissipation by
shocks (red) are plotted.  These quantities are integrated during the evolution,
so they represent the actual heating and cooling in these disks.
  For both resolutions of the standard case, the shock heating clearly
dominates over the cooling; the disk is heating up over the entire
evolution.  Energy is transported inefficiently for the highly optically thick
disk, and the disk evolves like an adiabatic simulation.  When the opacity
is reduced by a factor of 100, the cooling becomes much more important than it is in
the standard simulation, with radiative energy losses becoming comparable to heating
by shocks. In addition, the shock heating rate becomes somewhat larger during the burst. 
The 1024 standard and $10^{-2}$ $\kappa$ show additional heating.  This heating is quite strong in the 1024 standard simulation, and it must be stopped even earlier than the 512 standard due to high temperatures, as discussed above.  The 1024 $10^{-2}$ $\kappa$ is very similar to its 512 counterpart.  There is additional heating and cooling during the burst, but the heating and cooling curves for both resolutions roughly track each other.   The adiabatic curve, which is only shown to 33 yr (1 orp), is difficult to see because it tracks the standard cases extremely closely. 

When the opacity is reduced by a factor of $10^3$, the
radiative cooling is even more efficient and surpasses shock heating. The $10^{4}$ opacity reduction shows the strongest
shock heating and the fastest disk cooling.  The 1024 $10^{-4}$ $\kappa$ simulation curves track the 512 curves, with the burst contributing the most to the offset of the curves.  Unlike the higher opacity simulations, for $10^{-4}~\kappa$, there is less shock heating and less total cooling at higher resolution.  The 1024 $10^{-3}$ $\kappa$ simulation  roughly tracks the 512 counterpart, with the strongest deviation near 66 yr (2 orp).
The adiabatic curve is shown again as a reference. As one would expect from analytic arguments, the opacity has a profound effect on
disk cooling and, consequently, on spiral shocks.

The effect of opacity on energy losses is also demonstrated in
Figures 11 and 12.  In these figures, brightness temperature maps 
for the 1024 simulations are shown for similar times as in Figure 6,  where $T_b=(\pi
I_+/\sigma)^{1/4}$ and $I_+$ is vertical outward intensity as calculated by the ray solution in
CHYMERA.  In addition, we
define a cooling temperature $T_c=(\int_0^{\infty}\left[-\nabla\cdot{\bf F}\right]dz/\sigma)^{1/4}$, which is the temperature
that corresponds to a given column's effective flux if all of the energy were to
leave the column vertically.  If the column is being heated, $T_c$ is set to zero and appears as black in the images.

As the opacity is lowered, the spiral structure becomes more clearly outlined, and 
the disk becomes brighter because the photons can leave from hotter
regions.  The brightness maps also demonstrate that sustained fast cooling by convection 
is absent.  If hot gas near an optically thick midplane is quickly transported to altitudes where $\tau\sim1$, convection can, in principle, enhance disk cooling.  However, the efficacy of convective cooling is controlled by the rate at which that energy can be radiated from the photosphere of the disk.  If localized convective flows were responsible for fast cooling in the optically thick disks, one would expect to find strong, localized $T_b$ and $T_c$ enhancements.  
For example, the standard and $10^{-2}~\kappa$ simulations would need to have regions as bright as the $10^{-4}~\kappa$ simulation.    There are thin regions along the spiral shocks in the $10^{-2}~\kappa$ simulation that show large $T_c$, but they do not correspond to enhancements in $T_b$ as well.  This suggests that the energy from the spiral shocks is not transported efficiently to altitudes where it can be quickly lost from the disk.  This assessment is supported by the net radiative heating in regions surrounding the strong shocks, e.g., the black outlines (areas experiencing net radiative heating) around the thin, hot spiral waves.  The corresponding energy flux that is needed for convection to be sustained and to cool the disk does not occur (see also Boley et al.~2006, 2007b, Nelson 2006, \& Rafikov 2007).

The importance of the opacity for disk cooling is also shown in Figures 13 through 15.
On these plots, three quantites are shown: the Toomre $Q$, the mass-weighted
$\Gamma_1$, and the $\zeta=t_{\rm cool}\Omega/f(\Gamma_1)$ profiles. For the $t_{\rm cool}$ curves, the
cooling times are calculated by dividing the azimuthally and vertically
integrated internal energy by the azimuthally and vertically integrated radiative cooling rate for
each annulus in the disk.
The $f(\Gamma_1)$ is the critical value of $t_{\rm cool}\Omega$, below which fragmentation is expected (Gammie 2001), for the corresponding $\Gamma_1$.  Rice et al.~(2005) demonstrated that
$f(7/5)\approx12$ and $f(5/3)\approx6$.  Based on these values, we assume $f(\Gamma_1)\approx-23\Gamma_1+
44$ for the stability analysis presented here. One should keep in mind that $f(\Gamma_1)$ is not a strict threshold and that the relation is approximate, especially for simulations that permit the cooling time to evolve with the disk (Johnson \& Gammie 2003; Clarke et al.~2008).  Regardless, it serves as a general indicator for disk fragmentaiton.

As Figures 13 through 15 demonstrate, the
disk is approaching a state of constant $Q$ for a wide range of radii, as expected in an asymptotic phase (Mej\'ia et al.~2005).  The
mass-weighted average $Q$s between 3 and 6 AU are given in Table 1 and tend to decrease, as expected, when the opacity decreases.  They do not seem to be greatly affected by resolution. The
variation in the average $Q$ with opacity is roughly consistent with the $A_+$ measurements.  One should keep in mind that the $Q$ profiles represent snapshots of the disk, while the $A_+$ measurements are time averages.  The standard simulation
is not shown in the stability plots, but the analysis of the 512 standard disk shows that it is highly stable against fragmentation.

Because the cooling times fluctuate rapidly, we average the integrated cooling rates and
the internal energies for the, roughly,
65- and 70/80-year snapshots.  The cooling time profiles for the appropriate $\Gamma_1$
only drop substantially below unity over extended regions where $Q$ is high; these
disks are stable against fragmentation.  As the opacity is lowered, the cooling times decrease
as well, with the $10^{-4}$ $\kappa$ simulation close to the fragmentation limit.
It should also be noted that the only disk that behaves like a constant $\Gamma_1$ disk
is the standard opacity simulation (not shown), and for all other simulations, a constant $\Gamma_1$ is inappropriate (see Boley et al.~2007a).

Mass in these disks is redistributed efficiently during the activation of GIs.
Time-averaged mass fluxes are calculated by differencing the mass inside a cylinder at two different times.  This represents the average mass flux as calculated by the second-order hydrodynamics scheme, and it is independent from the fluid element tracer.  
The inward and outward accretion rates vary, and in all simulations, can be well above $10^{-4}~M_{\odot}$ yr$^{-1}$ (Fig.~16).

\subsection{Shocks}

 Due to our resolution, shocks are spread over scales of about $10^{12}$ cm or larger, whereas 1D calculations of chondrule-producing shocks give structures $
\sim 10^{10}$ cm (Desch \& Connolly 2002).   To overcome this difficulty, we measure the
pressure difference between the pre- and post-shock regions to calculate the Mach number $\mathcal{M}$.
 Using the pressure and temperature histories for each fluid element, a possible shock is identified whenever the $d T/d t$ changes from negative to positive then back to negative. The pre-shock flow is taken to be the first sign switch, and the post-shock flow is the second sign switch. Let $\eta=\left(p_2-p_1\right)/p_1$ be the fractional pressure change, where $p_1$ and $p_2$ are the pre- and post-shock pressures, respectively, and where $\gamma$ is the average of the pre- and post-shock first adiabatic index.    If $\mathcal{M}^2=\left(\gamma+1\right)\eta/\left( 2\gamma\right) + 1\ge 2$, the event is counted as a shock.  Once $\mathcal{M}$ is determined, the pre-shock velocity $u_1=\mathcal{M}c_{s1}$ is calculated, where $c_{s1}$ is the pre-shock adiabatic sound speed of the gas.  The shock strengths are also derived using the ratio of the post-shock to pre-shock temperatures:
 $T_2/T_1=\left(2\gamma\mathcal{M}^2-\left(\gamma-1\right)\right)\left(\left(\gamma-1\right)\mathcal{M}^2+2\right)/\left( \left( \gamma+1\right)^2\mathcal{M}^2\right)$.
Both approaches should give the same answer whenever the shocks are cleanly defined and adiabatic.  However, If radiative cooling becomes very efficient, then assuming adiabatic shock conditions may be incorrect.  Furthermore, secondary waves or shocks (Boley \& Durisen 2006) may make identifying pre- and post-shcok regions very difficult.  Neither method is clearly favored, and both will only provide crude estimates.

We estimate a shock to be
chondrule-forming when $u_1$ lies in a 1 km s$^{-1}$ band
between 5 km s$^{-1}$ and 11 km s$^{-1}$ and when the pre-shock density is between $\log
\rho(\rm~g~cm^{-3})$ = -8.5 and -10.5.  This estimate is based on the Desch \& Connolly (2002) 1D  calculations,  similar to the relation used to locate chondrule-producing shocks in Figure 1.  
Chondrule-forming shocks should be thought of as having the potential to form
chondrules, but may not yield chondritic material due to
incompatible dust to gas ratios, too high or low cooling rates, and fractionation.

 Table 2 indicates shock information as extracted from the pressure histories of the fluid elements. Estimating shock strengths with the temperature ratio yields comparable numbers except for the chondrule-forming shocks (see below). Column two shows approximately the
total number of detected shocks (TS) with $\mathcal{M}^2\ge2$, and column three displays
the average number of shocks per fluid element.  Columns four (TS Mass) and five (CS Mass) show the 
total dust mass (see below) that goes through a shock with $\mathcal{M}^2\ge 2$ and the total
dust mass that encounters a chondrule-forming shock, respectively.  To estimate how much dust is processed in shocks, we assign each fluid element a mass by calculating the total
disk mass within some $\Delta r$ and distributing that material evenly among all fluid elements in that $\Delta r$.  
A gas to solids ratio of 100 is assumed everywhere
for the dust processing calculations.  Although this is inconsistent with the growth and settling of solids assumed in six of the simulations, we do not worry about this detail inasmuch
as the midplane will process more solids per shock and the high-altitude shocks
will process less.  Finally, column
six shows the total number (Total FE) of fluid elements that remain after the 70 yr time period, i.e., the elements that
were not accreted onto the star or fluxed into background density regions.

We check whether the total number of detected shocks  is reasonable by evaluating the expected number of shocks in a Keplerian disk with $m$-arm spiral waves.   If the corotation
radius for the pattern is $r_p$, the number of shocks 
during some time period $\Delta t$ for a fluid element orbiting at $r$ is 
\begin{equation}N_S=\frac{m (GM_{\rm star})^{1/2}}{2\pi r^{3/2}}\bigg | 1-
\left(r/r_p\right)^{3/2}\bigg | \Delta t.\end{equation}
Evaluating equation (6) for 1000 fluid elements
evenly distributed in annuli between 2 and 8 AU, with $r_p=4$ AU,  with $m=3$, and with $\Delta t=60$ yr yields approximately
8000 shocks.  So the number of detected shocks in these simulations is reasonable. 

The total number of shocks per fluid element are roughly consistent for each simulation except for the $10^{-4}~\kappa$ runs. The more flocculent spiral structure  in these simulations not only produces a larger number of shocks in the disk, but also creates candidate chondrule-producing shocks for four fluid elements; the 1024 $10^{-3}~\kappa$ simulation has one fluid element that experiences a possible chondrule-forming shock.   Several thousand $M_{\oplus}$ of dust are pushed through shocks in each simulation, implying that almost all dust experiences a shock with $\mathcal{M}^2\ge2$. In addition, a few $M_{\oplus}$ of dust experience chondrule-forming shocks in the $10^{-4}~\kappa$ runs.  We remind the reader that these are only crude, order of magnitude estimates. When the temperature ratio is used, shocks with chondrule-forming strengths are {\it undetected}, but the total dust mass pushed through shocks remains the same to about one percent.

\section{Discussion and Conclusions}

In this section, we review the implications of the results of this study for disk fragmentation, the effects of opacity on disk cooling, the FU Ori phenomenon, and chondrule formation.  We remind the reader that the simulations presented here are meant to be a numerical experiment that explores the possible connection between chondrules, FU Ori outbursts, and bursts of GI-activity.  Moreover, this experiment provides a systematic study of the effects of opacity on disk cooling, which complements the Cai et al.~(2006) metallicity study and provides a test bed for disk fragmentation criteria.

\subsection{Fragmentation}

 After the onset of the burst, none of the disks fragments, and only
the 512 $10^{-4}$ $\kappa$ simulation shows dense knot formation during the peak of the burst (10-30 yr).  These knots do not break from the spiral wave even in the 1024 simulation, and so clump formation does not seem to be missed due to poor resolution. One should also keep in mind that for these simulations, the disk is first relaxed to equilibrium with standard opacity and then the opacity is suddenly dropped by a factor of 10$^{4}$ at $t=0$.  In effect, the dust settling is treated as instantaneous.  Knot formation may not occur in a disk with more realistic settling timescales.  As discussed below, this is also a caveat for the chondrule formation results.  On the other hand, it does suggest that disk fragmentation might occur inside 10 AU under even more extreme, but perhaps unphysical, conditions than we have modeled.  

The stability of these disks against fragmentation is supported by Figures 13 through 15.  The cooling rates for the standard and $10^{-2}~\kappa$ simulations are too low to cause disk fragmentation.  The $10^{-3}$ and  $10^{-4}~\kappa$ simulations do have areas where $\zeta = t_{\rm cool}\Omega/f(\Gamma_1) \lesssim 1$, but the disks approach stability against GIs in those regions ($Q\gtrsim1.7$; Durisen et al.~2007).   As noted in \S 4.2, $\zeta \lesssim 1$ is not a strict instability criterion, but it does serve as an estimate for disk stability against fragmentation.  For no simulation is $\zeta$ well below unity where $Q$ is also $\lesssim1.7$.

Lowering the opacity increases the cooling rates.  The $10^{-4}$ $\kappa$ simulations exhibit the most rapid cooling because the midplane optical depths are near unity, which results in the most efficient radiative cooling possible. {\it Changes in dust opacity have a profound effect on disk cooling}.  Although not modeled, if the opacity were to continue to drop such that the midplane optical depth becomes well below unity, cooling would once again become inefficient.  In addition, supercooling of the high optical depth disks (standard and $10^{-2}~\kappa$) by convection does not occur.  Based on our results, we believe that hydraulic jumps as a result of shock bores (Boley \& Durisen 2006) rather than convection are a better explanation for the upwellings around spiral arms reported by Boss (2004a) and Mayer et al.~(2007).   These findings are consistent with analytic arguments by Rafikov (2005, 2007), with numerical studies of disk fragmentation criteria by Gammie (2001), Johnson \& Gammie (2003), and Rice et al.~(2005), and with global disk simulations where tests of the radiation algorithm and/or careful monitoring of radiative losses were performed (Nelson et al.~2000; Boley et al.~2006, 2007b; Stamatellos \& Whitworth 2008).  

The radiative algorithm used in CHYMERA has passed a series of radiative transfer tests that are relevant for disk studies, including permitting convection when expected.  Our conclusions regarding fragmentation are based on mulitple analyses: surface density rendering (Figures 4-7), an energy budget analysis (Figures 9 and 10), cooling temperature and brightness maps (Figures 11-12 ), and a cooling time stability analysis (Figures 13-15).   It is also important to point out that these results do not contradict Stamatellos et al.~(2007) or Krumholz et al.~(2007), who find fragmentation in massive, extended disks ($>100$ AU).

It is also pertinent to demonstrate that the CHYMERA code can detect fragmentation when cooling rates are high and $Q$ is low.  Figure 17 shows a snapshot for a simulation similar to the 512 $10^{-4}~\kappa$ simulation, but with the divergence of the fluxes artificially increased by a factor of two.  In the normal simulation, kinks in the spiral waves form during the onset of the burst.  
As discussed above, the disk is very close to the fragmentation limit, but the knots do not break from the spiral wave.  Figure 15 suggests, although for a later time, that increasing the cooling rates by a factor of two would drop $\zeta$ well below unity in low-$Q$ regions. Figure 17 confirms that the disk fragments with such enhanced cooling.   However, we remind the reader that the $10^{-4}~\kappa$ simulation has the optimal optical depth for cooling ($\tau\sim1$).  We cannot imagine any physical process that could cause a factor of two enhancement in cooling.

Three clumps form, one for each spiral wave, between 4 and 5 AU.  The location of clump formation is consistent with the prediction by Durisen et al.~(2008) that a spiral wave is most susceptible to fragmentation near corotation.  One of the clumps survives for several orbits and eventually passes through the inner disk boundary.  Resolution is always a concern for simulations.  Because these results are consistent with analytic fragmentation limits and numerical fragmentation experiments, we conclude that CHYMERA can detect fragmentation at the resolutions employed for this study. 

As discussed above, when the opacity is abruptly decreased by a factor of 10${^4}$, the disk does approach fragmentation-like behavior.  A similar effect is reported by Mayer et al.~(2007), who find that their disk only fragments when the mean molecular weight is suddenly switched from $\mu=2.4$ to $\mu=2.7$.  However, the simulation presented by Boley et al.~(2006) was accidentally run with $\mu=2.7$, and Cai et al.~(2006, 2008) purposefully ran their simulations at the high $\mu$ for comparison with the Boley et al.~results.  None of the studies reported disk fragmentation.  This may indicate that when a disk approaches fragmentation shortly after a sudden switch in a numerical parameter, e.g., opacity here and $\mu$ in Mayer et al., the fragmentation may be numerically driven rather than physically, especially because sudden changes in cooling rates make disks more susceptible to fragmentation (Clarke et al.~2008). Regardless, the results do indicate that disk fragmentation by GIs may be possible under very extreme conditions. Whether such conditions are physically possible or realistic remains to be shown.  Based on our results here and in earlier papers, disk fragmentation for $r \lesssim 10$s of AU appears to be a yet-unproven exception rather than the norm.

\subsection{FU Ori Outbursts}

In these simulations, the strong bursts of GI activity provide high mass fluxes  ($\dot{M}\gtrsim10^{-4}~M_{\odot}$ yr$^{-1}$, see Fig.~16) throughout each disk.  Even though corotation
is at $r\sim4$ AU for the major spiral arms, the 2 AU region of the disk is
strongly heated.  Fluid elements approach peak temperatures of $T\sim1000$ K in all simulations (Fig.~18). It is not difficult to speculate that if a larger extent of
the disk were modeled, the temperatures due to shocks would be large enough
to ionize alkalis thermally and possibly sublimate dust (1400-1700 K).  If such a condition is met, then an MRI could activate and rapidly carry mass into the innermost regions of the disk.
 From these simulations it appears
to be plausible, but by no means proven, that a burst of GI activity as far out as 4 AU can
drive mass into the inner regions of the disk and create strong temperature fluctuations, which may then be responsible for a thermal instablility.  Although we are simulating very
massive disks, we note that such systems may exist during the Class I YSO phase.  Additional studies need to be conducted, preferably with a self-consistent buildup of a dead zone, to address the efficiency of this mechanism in low mass disks.

There are at least three observable signatures for this mechanism.  First, if a
GI-bursting mass concentration at a few AU ultimately results in an FU Ori phenomenon, then
one would expect to see an infrared precursor from the GI burst, with a
rise time of approximately tens of years.  Second, one would also
expect for a large abundance of molecular species, which would normally be
frozen on dust grains, to be present in the gas phase
during the infrared burst due to shock heating
(Hartquist 2007, private communication).  Third, approximately the first
ten AU of the disk should have large mass flows if the burst takes place near
$r\sim4$ to 5 AU.  We speculate based on these results that if the burst were to take place
at 1 AU, then high outward mass fluxes should be observable out to a few AU.

\subsection{Dust Processing} 
Each simulation shows that a large fraction of material
goes through shocks with $\mathcal{M}^2\ge2$. Although these shocks are weak,
their abundant numbers may result in the processing of dust to some
degree everywhere in the 2-10 AU region. In fact, such processing may be
necessary for prepping chondritic precursors for strong-shock survival  (Ciesla 2007, private communication). 

Based on the arguments presented in \S 2, the intent was to produce spirals with large
pitch angles by constructing a disk biased toward a strong, sudden GI-activation near 5 AU.  Even with this bias, the pitch angles of the spirals remain small, with $i\approx10^{\circ}$.  
Why are the pitch angles so small?  According to the WKB approximation, $\cot i = \mid k_r r/m\mid$ (Binney \& Tremaine 1987), where $k_r$ is the radial wavenumber for some $m$-arm spiral.  The most unstable wavelength for axisymmetric waves (Binney \& Tremaine 1987) is roughly $\lambda_u\approx 2\pi c_s/Q\kappa\approx2\pi h/Q$ for disk scale height $h$.   For a disk unstable to nonaxisymmetric modes, $\lambda_u$ corresponds to some $m$-arm spiral (e.g., Durisen et al.~2008).  By relating $k_r=2\pi\beta m/\lambda_u$, 
\begin{equation}
\cot i\approx \beta Qr/h
\end{equation} 
in the linear WKB limit ($\mid k_r r/m\mid \gg 1$), where $\beta$ is a factor of order unity that relates $\lambda_u$ to the $m$-arm spiral.  Because $\beta Qr/h \sim10$  in gravitationally unstable disks, the linear WKB analysis may be marginally applicable.  Equation (7) predicts that linear spiral waves in these disks should have pitch angles $i\approx6$ to 11$^{\circ}$ for $\beta$ between 1 and 1/2, respectively.  It appears that this estimate for $i$ extends accurately to the nonlinear regime, a result we did not expect.

The GI spirals are efficient at heating the disk and transporting angular momentum, but {\it not} at producing chondrules.  Only for the $10^{-4}$ $\kappa$ simulation are multiple candidate chondrule-producing shocks detected, and there is only one chondrule-producing shock in the 1024 $10^{-3}$ $\kappa$ simulation. These low opacity disks have relatively flocculent spiral morphologies, including kinks in spiral waves.  The $10^{-4}~\kappa$ simulations are on the verge of fragmentation. When interpreting these results, it is important to remember that the opacities were lowered abruptly and in a quite unphysical way. In addition, these detections are based on the fractional pressure change $\eta$.  If the temperature change is used, no chondrule-producing shocks are detected.  This discrepancy may be due to efficient radiative cooling, which may make the assumption of adiabatic shock conditions incorrect, and/or to confusion with additional waves induced by shock bores.  For the rest of this section, we assume that the pressure difference is the reliable shock identifier in order to discuss the implications of detecting chondrule-forming shocks in these disks.  A thermal and spatial history for a fluid element that experiences a possible chondrule-forming shock is shown in Figure 19. 

To estimate the occurrence of a chondrule-forming shock, we employ a generous
$u_1$-$\rho$ criterion (Fig.~20).  We do not take into account the optical depth criterion of Miura \& Nakamoto (2006) on grounds that the large scale over which these shocks take place can allow for chondrules to equilibrate with their surroundings (Cuzzi \& Alexander 2006).
However, one should be aware that the optical depth criterion of Miura \& Nakamoto
will likely exclude all chondrule-producing shocks inside 4 AU in the $u_1$-$\rho$ plane for
these simulations.  All candidate chondrule-forming shocks occur between $r\sim3$ and 5 AU and at altitudes that are roughly less than a third of the gas scale height.  Because a large degree of settling is assumed, these shocks are located in regions that may be consistent with the dusty environments in which chondrules formed (Wood 1963; Krot et al.~2005). 

We estimate that  $\sim 1~M_{\oplus}$ of dust is processed 
through chondrule-forming shocks.  Because these shocks are limited to the onset of the GI burst,
a few$\times10~M_{\oplus}$ of dust would be processed in the $10^{-4}$ $\kappa$ disks if they went through about ten outbursts.  If more outbursts take place than those that lead to an FU Orionis event, then more chondritic material could be produced.  In order to produce these shocks, the disk was pushed toward fragmentation by suddenly dropping the opacity by a factor of $10^{4}$.  We conclude that for bursts of GIs near 4 or 5 AU to produce chondrules, the disk must be close to fragmentation.

\section{The Unified Theory}
The hypothesis behind the work summarized here is that bursts of GI activity, dead zones, the FU Ori phenomenon, and chondrule formation are linked.  The general picture is that mass builds up in a dead zone due to layered accretion and that GIs erupt once $Q$ becomes low.  This activation of the instability causes a sudden rise in the mass accretion rate that heats up the disk inside 1 AU to temperatures that can sustain thermal ionization and an MRI.  The MRI shortly thereafter activates the thermal instability (Bell \& Lin.~1994; Armitage et al.~2001; Zhu et al.~2007).  In addition, the strong shocks process dust and form chondritic material in the asteroid belt and at comet distances.

Given the results of our simulations, we think the Boss \& Durisen conjecture that dead zones bursting at $r\gtrsim$ 4-5 AU can produce chondrules is {\it unlikely}, unless the disks are on the verge of fragmentation.  Because the conditions for fragmentation inside $r\sim10$s of AU are very difficulty to achieve, we conclude that this is not viable as a general chondrule-formation mechanism.  As a result, we return to Figure 1.  Based on our simple, analytic argument in equation (1), chondrule formation can take place in the asteroid belt and at cometary distances for GI-bursting dead zones between about 1 and 3 AU, even with $i\approx10^{\circ}$.  As suggested by Wood (2005), chondritic parent bodies may be representative of temporal as well as spatial formation differences.   Bursts that occur roughly between 1 and 3 AU seem to be able to accommodate this scenario, and could drive the FU Ori phenomenon (Armitage et al.~2001).  As the disk evolves, the location of the dead zone can vary, with multiple bursts occurring at a few AU.  Accretion rates between $10^{-6}$ and $10^{-7}~M_{\odot}~\rm yr^{-1}$ could build up a 0.01 $M_{\odot}$ dead zone every $10^4$ to $10^5$ years, respectively.  Some of these bursts may drive the FU Ori phenomenon and produce chondrules, but others may only be responsible for chondrule-formation events.  In this scenario, the disk need not fragment, and there should be between about 10 and 100 separate chondrule-formation events.  Future work is required to ascertain the plausibility of this modified version of the unified theory.

We would like to thank F.~Ciesla, S.~Desch, and L.~Hartmann for fruitful discussions.  We would also like to thank the anonymous referee for comments and suggestions that helped improve this manuscript. A.C.B.'s contribution was supported by a NASA GSRP fellowship, and R.H.D.'s contribution was supported by NASA grant NNG05GN11G. This work was supported in part by the IU Astronomy Department IT facilities, by systems made available by the NASA Advanced Supercomputing Division at NASA Ames, by systems obtained by Indiana University through Shared University Research grants through IBM, Inc., to Indiana University, and by dedicated workstations provided to the Astronomy Department by IU's University Information Technology Services.

\newcommand{\chondrites}{2005, in ASP Conf.~Ser.~341, Chondrites and the protoplanetary disk}

\begin{table}
\begin{center}
\caption[Simulation information]{Simulation information.  The name of the simulation indicates the fraction of the standard opacity that is used during the evolution of the disk. $Q_{\rm av}$ is the average $Q$ between 3 and 6 AU for a snapshot near 70/80 yr, depending on the simulation (see text). The $A_+$ measurement is the sum of the time-averaged Fourier components between roughly 55 and 80 yr.  }
\begin{tabular}{c c c c c }\\ \hline
Sim.~Name & Resolution $r,~\phi,~z$ & Duration & $Q_{\rm av}$ & $\left<A_+\right>$ \\ \hline\hline
Adiabatic & 256, 512, 64 & 33 yr \\
512 Standard &  256, 512, 64 & 80 yr & 1.7 & 1.0 \\
1024 Standard & 512, 1024, 128 & 55 yr & -- &\\
512 $10^{-2}~\kappa$&  256, 512, 64  & 110 yr & 1.7 & 1.3  \\
1024 $10^{-2}~\kappa$& 512, 1024, 128 & 110 yr & 1.8 & 1.6   \\
512  $10^{-3}~\kappa$&  256, 512, 64 & 110 yr & 1.5 & 2.1  \\ 
1024  $10^{-3}~\kappa$& 512, 1024, 128 & 110 yr & 1.5 &  2.2 \\ 
512 $10^{-4}~\kappa$&  256, 512, 64  & 110 yr & 1.4 &  2.1 \\ 
1024 $10^{-4}~\kappa$& 512, 1024, 128 & 110 yr & 1.4 & 2.4 \\ \hline
\end{tabular}
\end{center}
\end{table}

\begin{table}
\begin{center}
\caption[Shock information]{Shock information for the time period between 10 and 70 yr.  All estimates are based on the fractional pressure change $\eta$ (see text).  TS indicates the total shocks encountered by all fluid elements. TS/FE gives the average number of shocks for a fluid element. TS Mass is a rough estimate of the total dust mass pushed through shocks in each disk, and CS Mass is the total dust mass that encounters a chondrule-forming shock.  Finally, Total FE indicates the number of fluid elements that remain in the simulated disk at the end of the 60 yr time period.  The 1024 standard simulation is omitted because it was stopped after 55 yr.}
\begin{tabular}{c c c c c c c}\\ \hline
Sim. & TS & $\frac{\rm TS}{\rm FE}$ & 
            TS Mass   (M$_{\oplus}$)& CS Mass (M$_{\oplus}$)   & Total FE\\ \hline\hline

512 standard & 8700  & 9 & 4(3) & 0 & 992  \\
512 $10^{-2}$ $\kappa$ & 8600 & 9  & 4(3) & 0 & 990 \\
1024 $10^{-2}$ $\kappa$ & 9500 & $10$ & 4(3) & 0 & 962 \\
512 $10^{-3}$ $\kappa$ & 8300 & 9 & 4(3) & 0 & 978 \\ 
1024 $10^{-3}$ $\kappa$ & 9500 & 10 & 4(3) & $<1$ & 964 \\ 
512 $10^{-4}$ $\kappa$ & 9300 & 10 & 4(3) & 3 & 939 \\ 
1024 $10^{-4}$ $\kappa$ & 11000 & 12 & 5(3) & 2 & 913 \\ \hline

\end{tabular}
\end{center}
\end{table}

\clearpage

\begin{figure}
\begin{center}
\includegraphics[width=16cm]{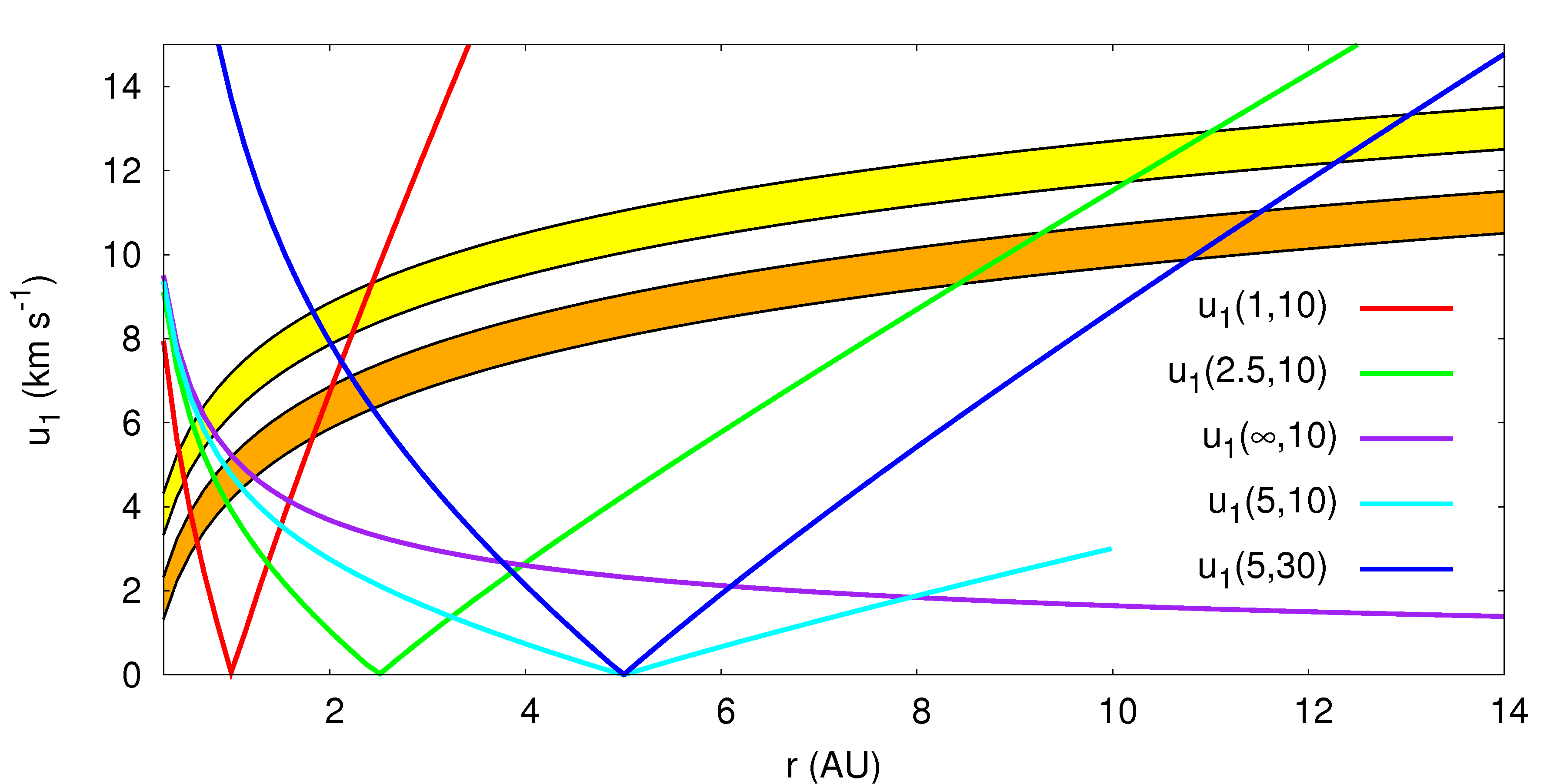}
\caption[Shock speed expectation]{Expected shock speeds $u_1$ based on equation (1). In the legend, \linebreak $u_1(1,10) = u_1(r_p=1{~\rm AU,}~i=10^{\circ})$ for corotation radius of the spiral pattern $r_p$ and pitch angle $i$. The colored regions highlight where chondrule formation is expected based on the Desch \& Connolly (2002) shock calculations (see text), with the yellow region appropriate for a MMSN density distribution and the orange for the same density distribution but with 10$\times$ the mass.  Shocks
that occur inside corotaton for $i=10^{\circ}$ will not produce chondrules between 1 and 5 AU.  
However, chondrules can be produced by these low pitch angle spirals in shocks outside corotation.
If the pitch angle is fairly open, such as $i\approx30^{\circ}$, then a spiral wave with a corotation near 5 AU can produce chondrules in the asteroid belt and at comet distances.}
\end{center}
\end{figure}

\clearpage

\begin{figure}
\begin{center}
\includegraphics[width=16cm]{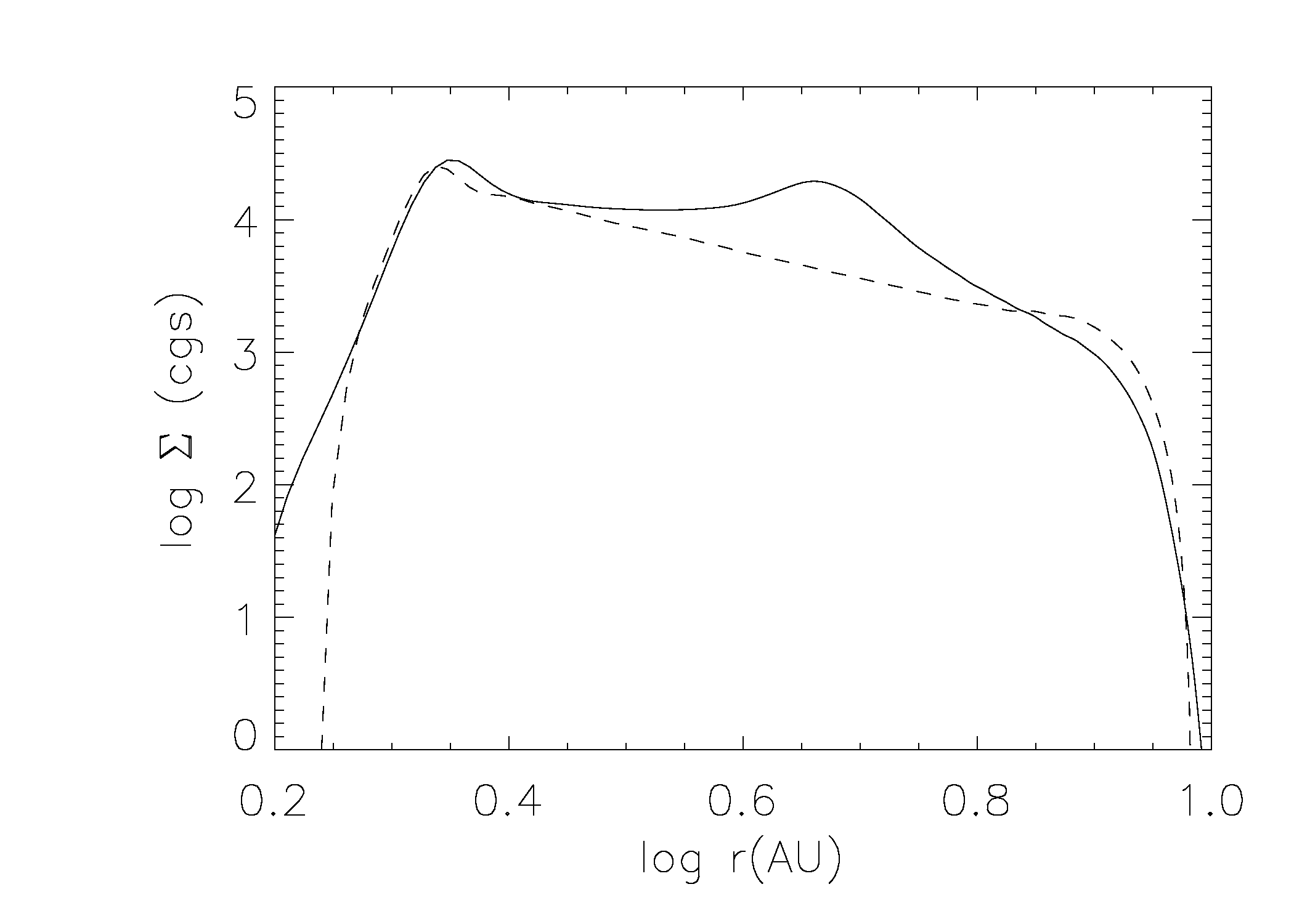}
\caption{Surface density profiles for the ICs with (solid) and without (dashed) the density enhancement.
The peak enhancement occurs near $\log (4.5~{\rm AU}) \approx 0.65$.}
\end{center}
\end{figure}

\clearpage

\begin{figure}
\begin{center}
\includegraphics[width=8cm]{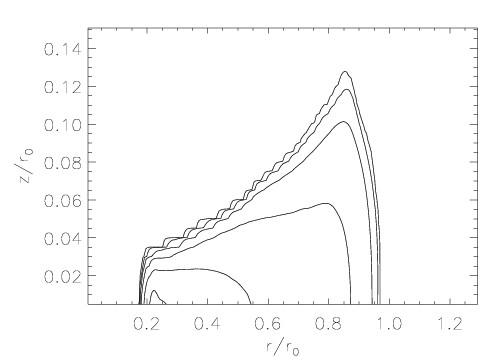}
\includegraphics[width=8cm]{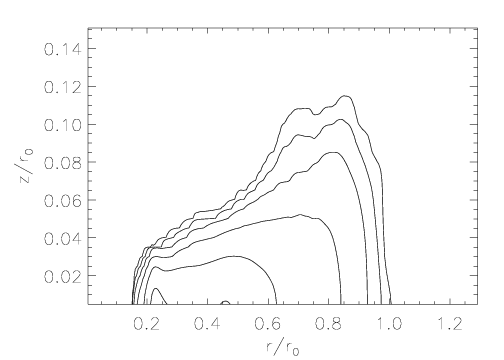}
\caption{Volume density contours for the initial model with (right) and without (left) the density enhancement. The normalization $r_0=10$ AU.}
\end{center}
\end{figure}

\clearpage

\begin{figure}
\begin{center}
\includegraphics[width=16cm]{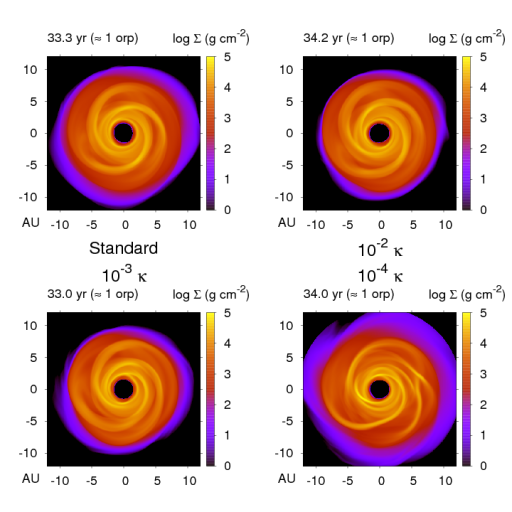}
\caption{Surface density plots for the 512 simulations around 33 yr.}
\end{center}
\end{figure}

\clearpage

\begin{figure}
\begin{center}
\includegraphics[width=16cm]{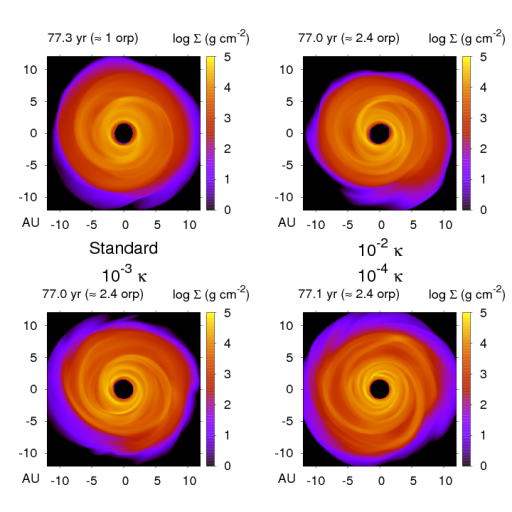}
\caption{Surface density plots for the 512 simulations around 77 yr.}
\end{center}
\end{figure}

\clearpage

\begin{figure}
\begin{center}
\includegraphics[width=16cm]{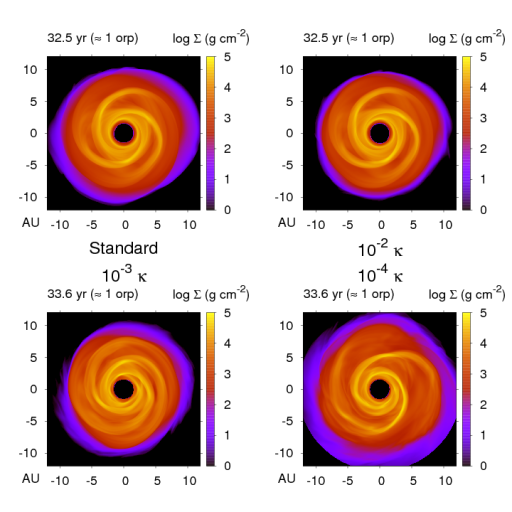}
\caption{Surface density plots for the 1024 simulations around 33 yr.}
\end{center}
\end{figure}

\clearpage

\begin{figure}
\begin{center}
\includegraphics[width=16cm]{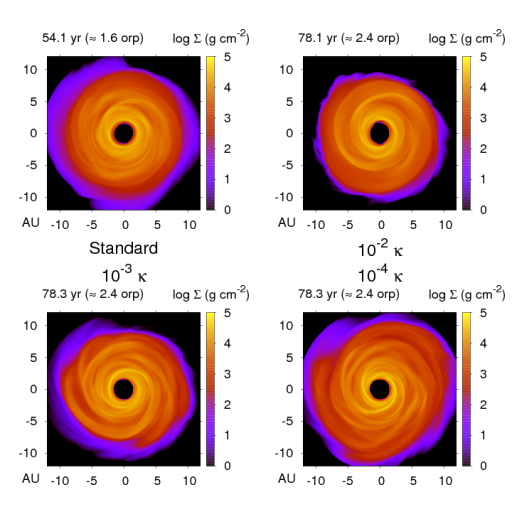}
\caption{Surface density plots for the 1024 simulations around 78 yr.  The 1024 standard simulation
is only run to about 55 yr (see text).}
\end{center}
\end{figure}

\clearpage

\begin{figure}
\begin{center}
\includegraphics[width=16cm]{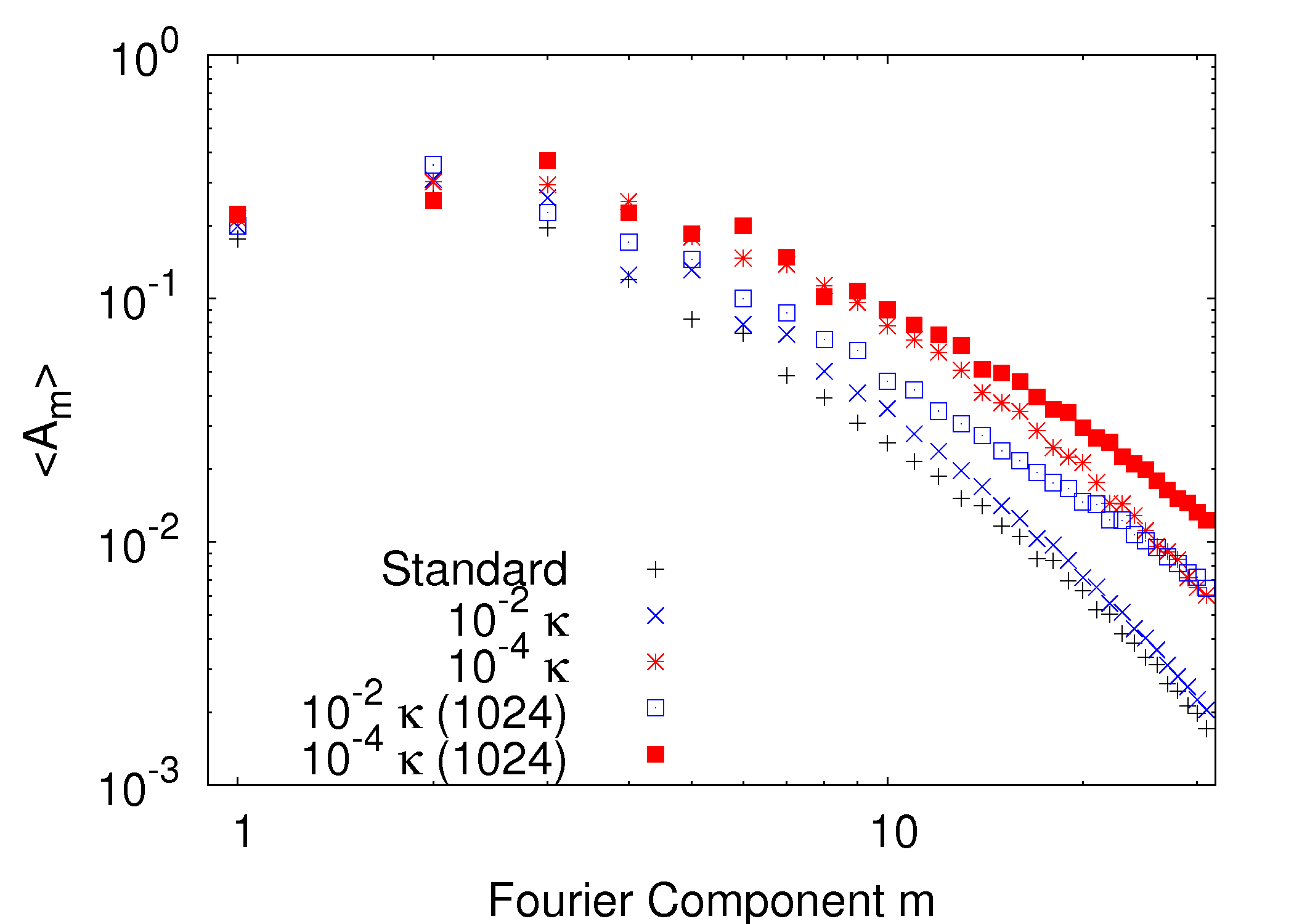}
\caption{$A_m$ profiles for the the 512 simulations (crosses) and the 1024 simulations (squares).  The standard simulation does not have a 1024 counterpart because it was stopped at about 55 yr (see text). These profiles demonstrate that the low-order structure dominates in the disk, with the Fourier components dropping off quickly near $m\sim10$. The lower opacity simulations have stronger $A_m$s, but there do seem to be some resolution effects present at large $m$.  The $10^{-3}~\kappa$ simulation is not shown for readability; it falls between the $10^{-2}$ and $10^{-4}~\kappa$ simulations and the 512 and 1024 resolutions follow each other closely.}
\end{center}
\end{figure}

\clearpage

\begin{figure}
\begin{center}
\includegraphics[width=16cm]{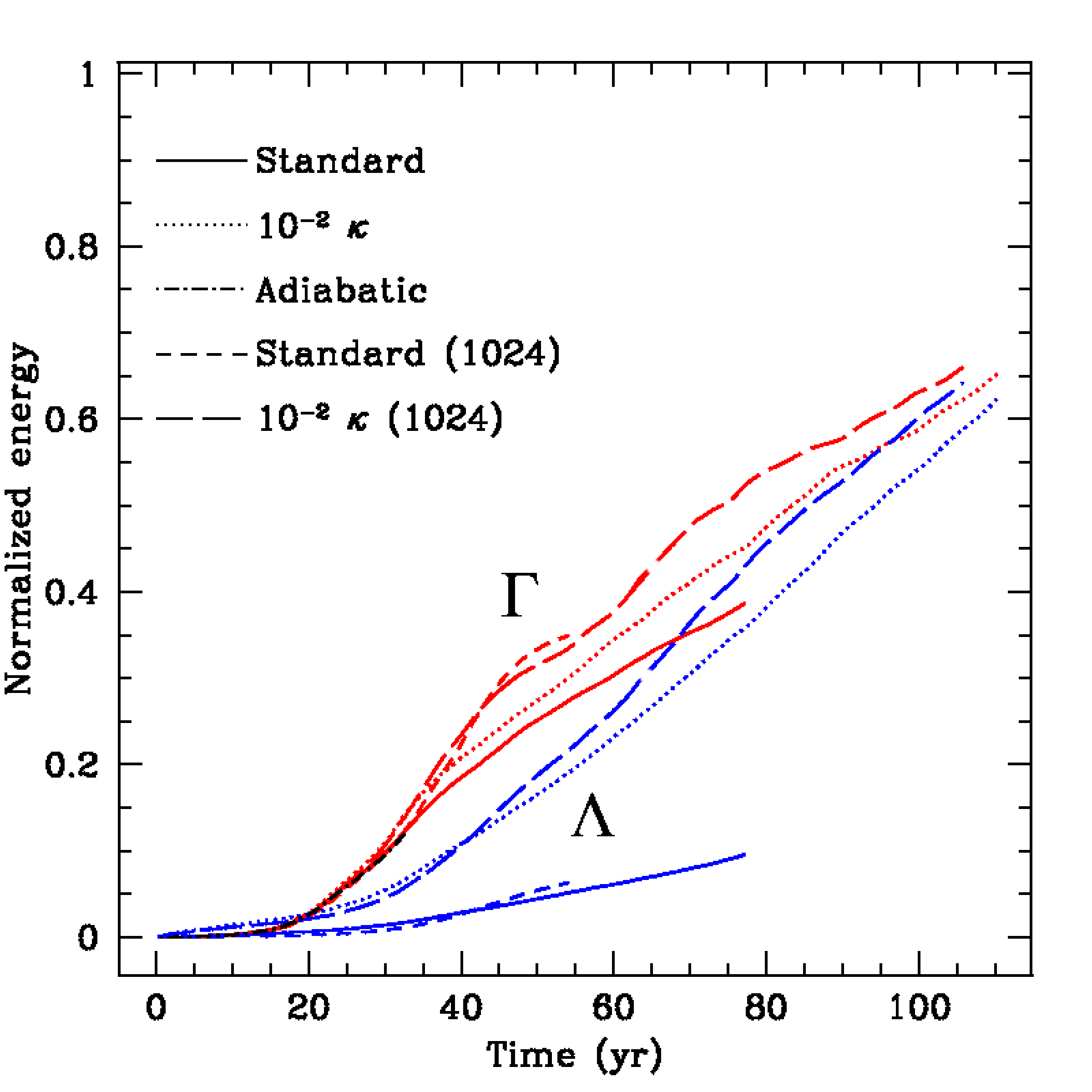}
\caption{Cumulative energy losses by radiation (blue) and heating by shocks (red) for the standard and $10^{-2}~\kappa$ simulations. The standard simulations are stopped early because the temperatures in the disk become too high to evolve due to inefficient cooling.  The shock heating in the adiabatic (no cooling) simulation is indicated by the dark, dot-dash line.  It is difficult to see because it follows the standard curves closely.  For clarification, a $\Lambda$ is shown adjacent to the cooling curves and a $\Gamma$ is shown next to the heating curves.}
\end{center}
\end{figure}

\clearpage

\begin{figure}
\begin{center}
\includegraphics[width=16cm]{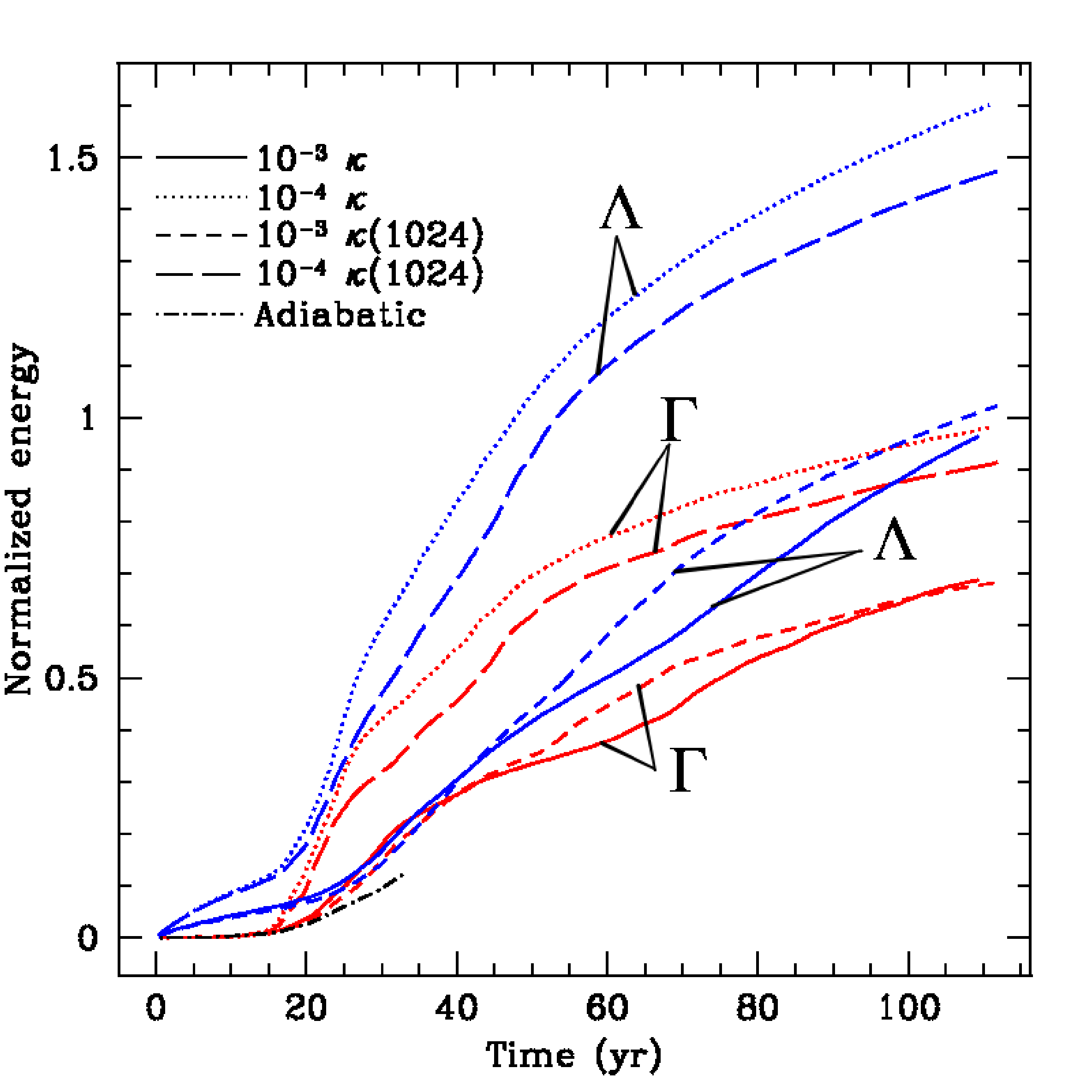}
\caption{Similar to Figure 9, but for the $10^{-3}$ and $10^{-4}~\kappa$ simulations. }
\end{center}
\end{figure}

\clearpage

\begin{figure}
\begin{center}
\includegraphics[width=16cm]{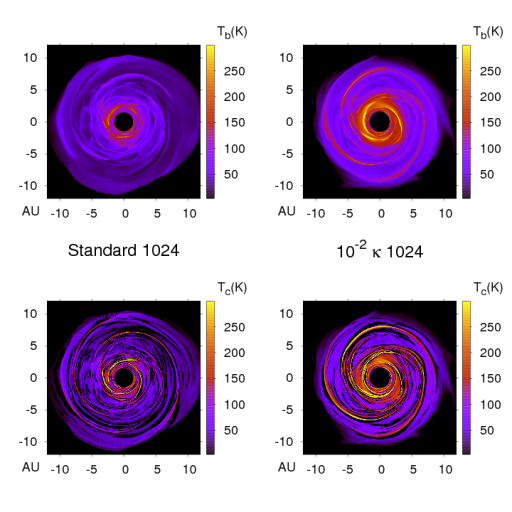}
\caption{$T_b$ and $T_c$ maps for the 1024 standard  and $10^{-2}$ $\kappa$ simulations. As the opacity is lowered, the disks become much more efficient at cooling.  See also Figure 12.}
\end{center}
\end{figure}

\clearpage

\begin{figure}
\begin{center}
\includegraphics[width=16cm]{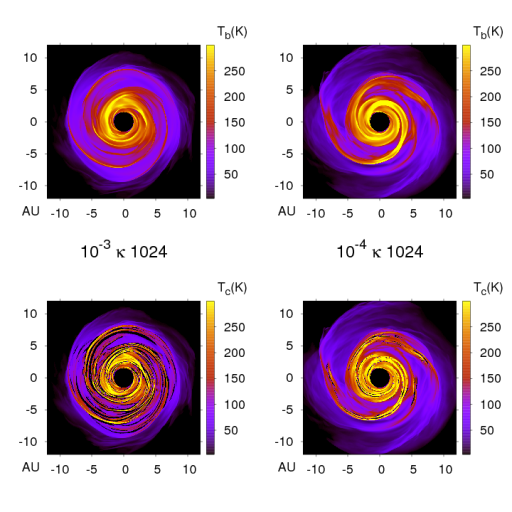}
\caption{Same as Figure 11, but for the 1024 $10^{-3}$ and $10^{-4}$ $\kappa$ simulations.  }
\end{center}
\end{figure}

\clearpage

\begin{figure}
\begin{center}
\includegraphics[width=12cm]{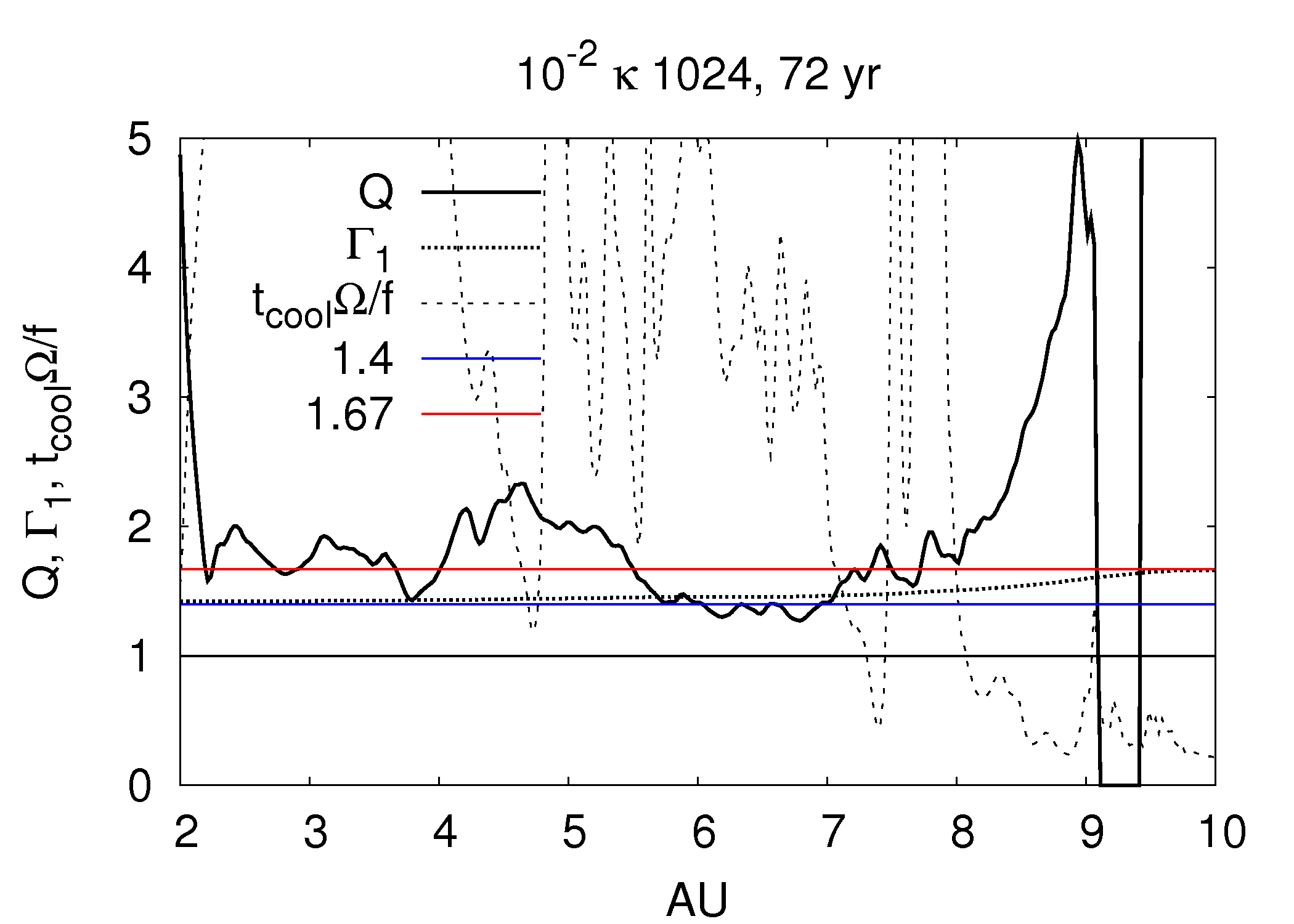}
\caption
{Stability assessment for the 1024 $10^{-2}~\kappa$ simulations. The ordinate shows values for $Q$, $
\Gamma_1$, and $t_{cool}\Omega/f(\Gamma_1)$ as a function of $r$, where 
$f(\Gamma_1)\approx-23\Gamma_1+44$.  If
$t_{cool}\Omega/f(\Gamma_1) > 1$, then the disk is 
stable against fragmentation.}
\end{center}
\end{figure}

\clearpage

\begin{figure}
\begin{center}
\includegraphics[width=12cm]{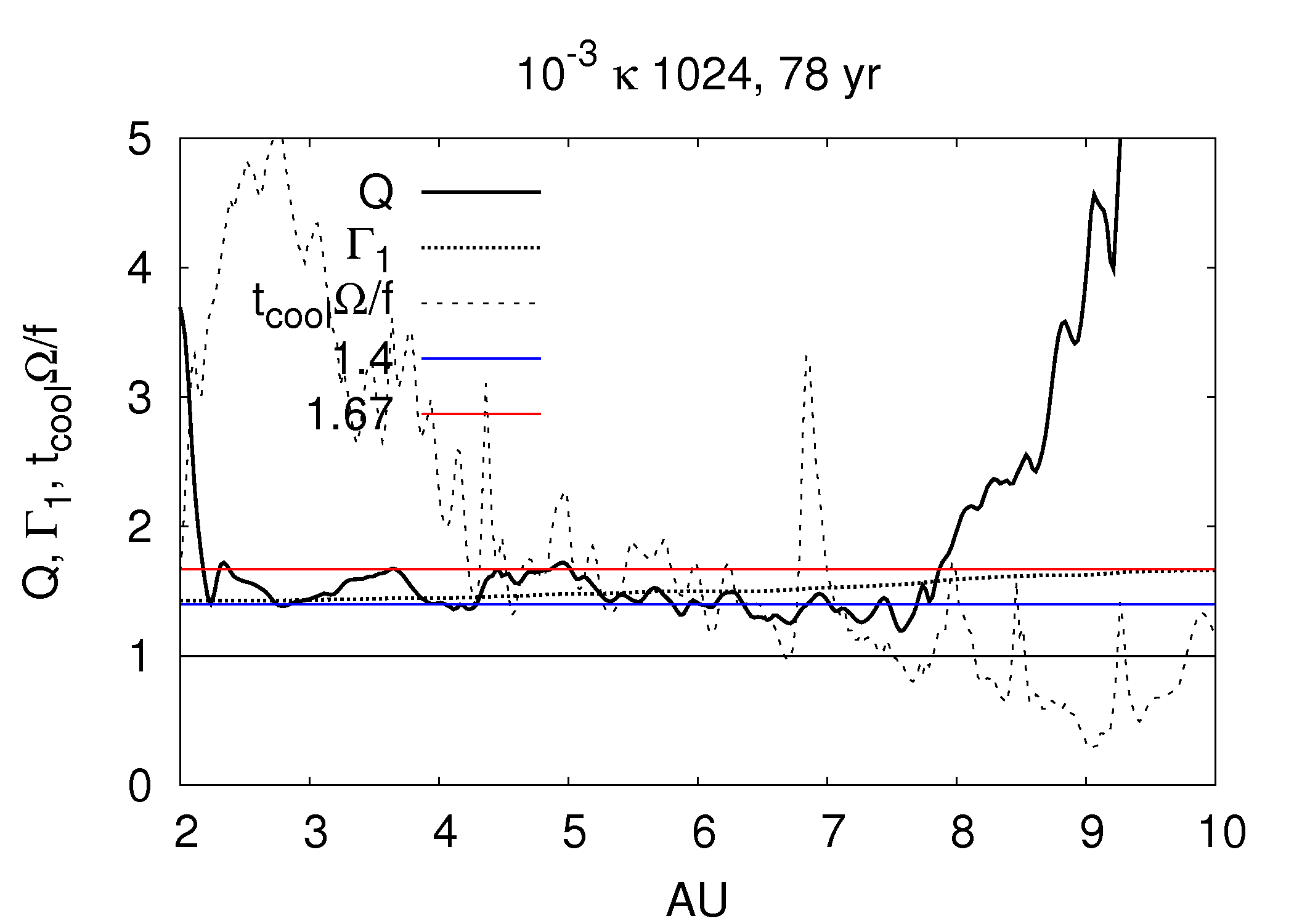}
\caption{Same as Figure 13, but for the 1024 $10^{-3}$ $\kappa$ simulation.}
\end{center}
\end{figure}

\clearpage

\begin{figure}
\begin{center}
\includegraphics[width=12cm]{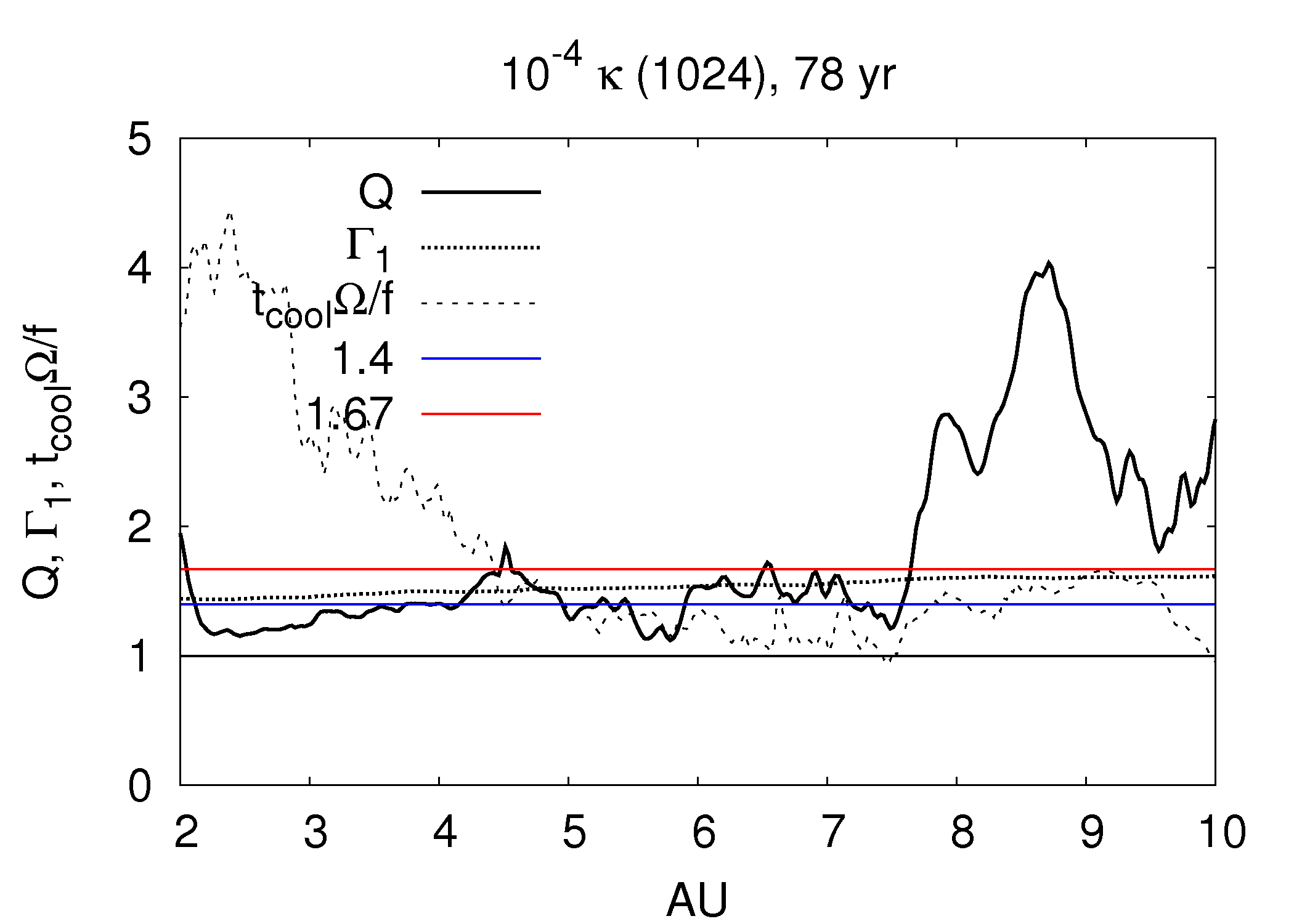}
\caption{Same as Figure 13, but for the 1024 $10^{-4}$ $\kappa$ simulation.}
\end{center}
\end{figure}

\clearpage

\begin{figure}
\begin{center}
\includegraphics[width=16cm]{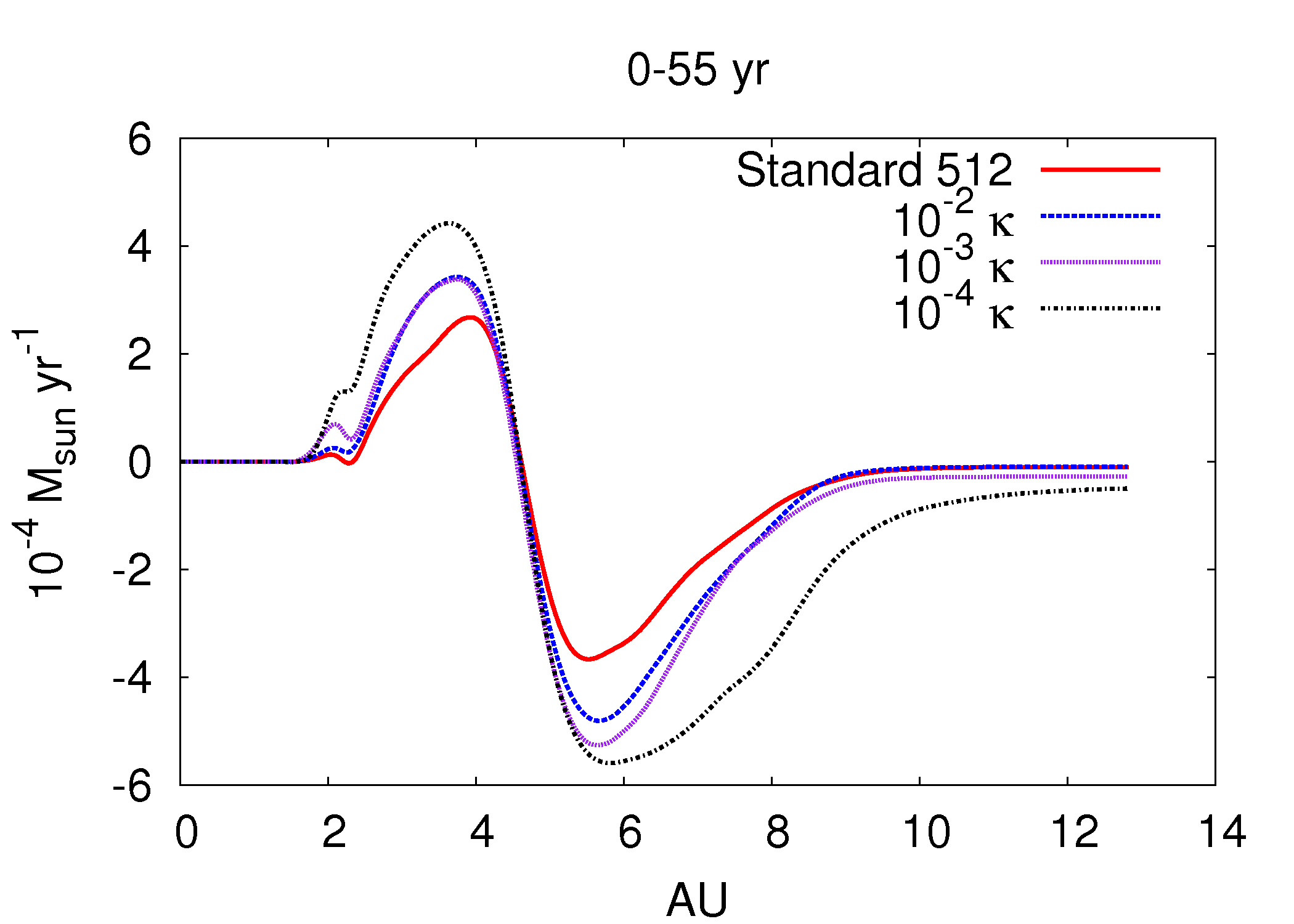}
\caption{Time-averaged mass fluxes during the 0-55 yr period for the 512 Standard and the 1024 reduced opacity simulations. 
For each simulation, the mass fluxes are at FU Ori outburst  levels. Positive values here correspond to net inflow, and negative values net outflow.}
\end{center}
\end{figure}

\clearpage

\begin{figure}
\begin{center}
\includegraphics[width=16cm]{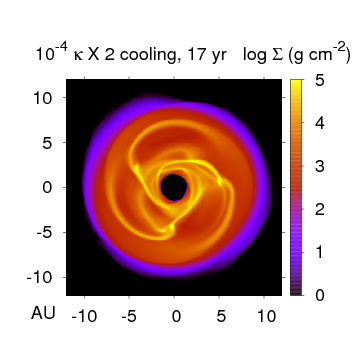}
\caption{Similar to the 512 $10^{-4}~\kappa$ simulation, but with the cooling
artificially enhanced by a factor of two.  The fragmentation is consistent with
analytic predictions.}
\end{center}
\end{figure}

\clearpage

\begin{figure}
\begin{center}
\includegraphics[width=3in]{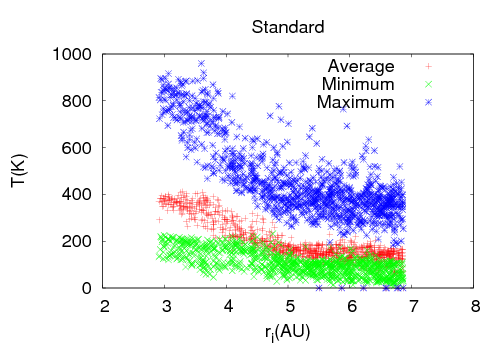}\includegraphics[width=3in]{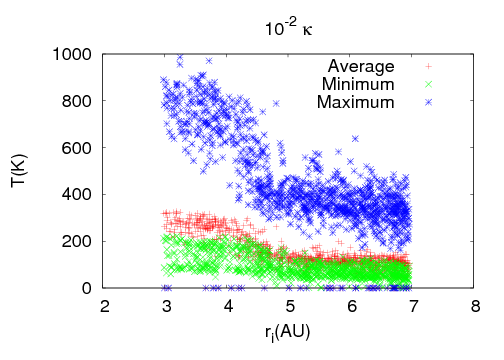}
\includegraphics[width=3in]{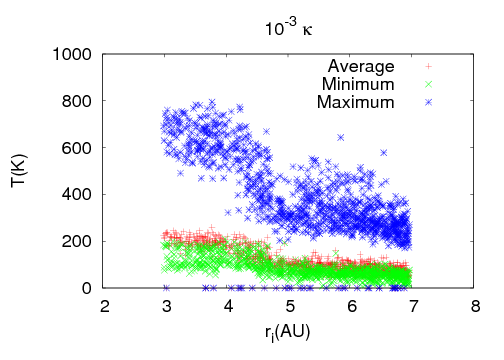}\includegraphics[width=3in]{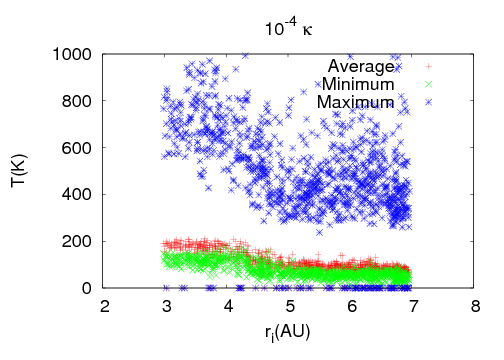}
\caption{Initial $r$ and minimum, average, and maximum temperature plots for fluid elements in the 512 Standard and the 1024 reduced opacity simulations.  The temperature variations are quite large, approaching maximum temperatures of 1000 K. The values on the abscissa are due to the fluid elements
that were lost from the simulated disk.}
\end{center}
\end{figure}  

\clearpage

\begin{figure}
\begin{center}
\includegraphics[width=5.95in]{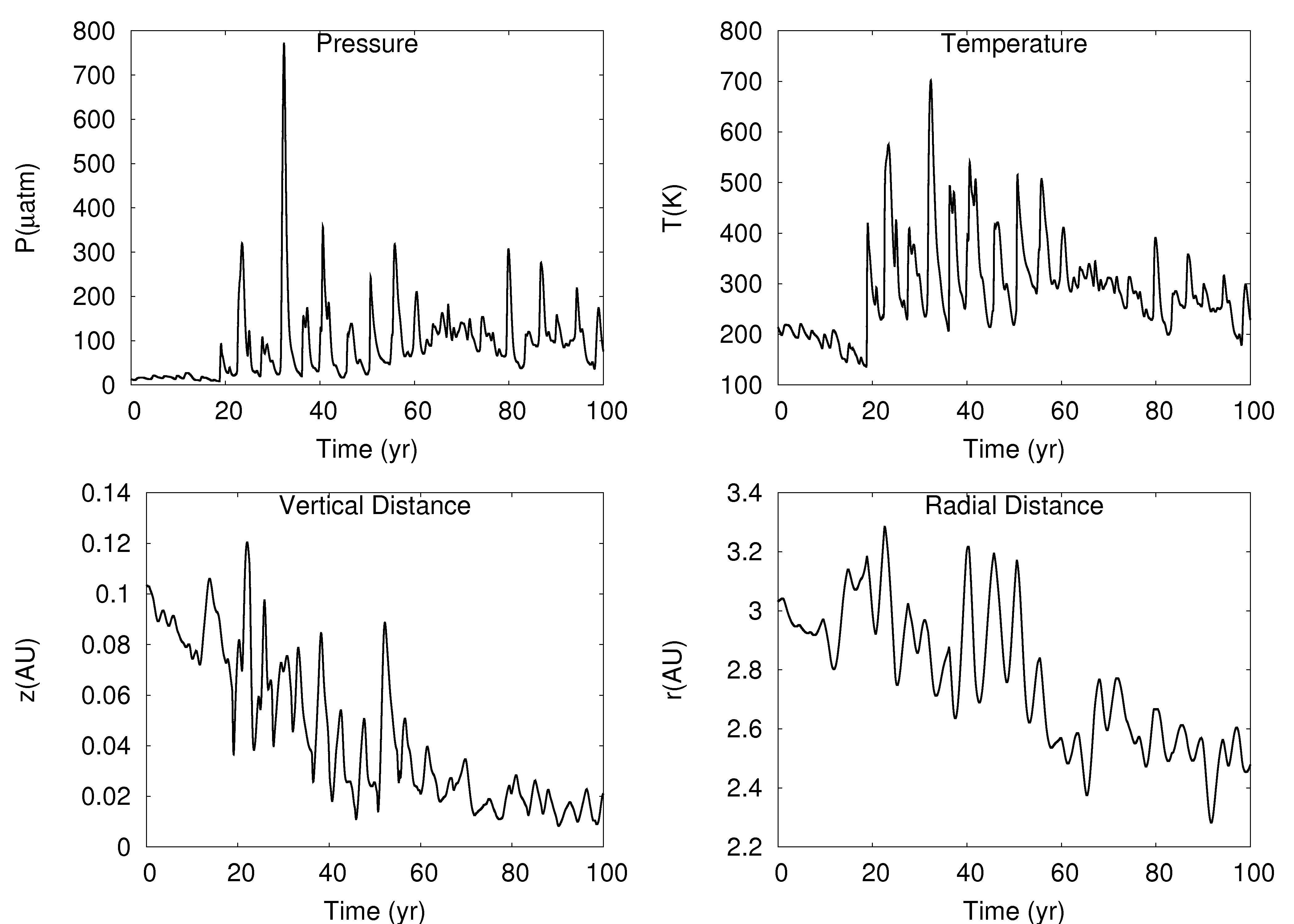}
\caption{A thermal and spatial history for a fluid element in the
the 1024 $10^{-4}~\kappa$ simulation.  The fluid element shows strong variations in
radius and height as it passes through shocks, with a net inflow due to gravitational 
torques.  Strong variations in the pressure and temperature are also present.  The strong
shock near 30 yr is a potential chondrue-forming shock
based on the pressure profile.  However, the temperature peak is not as high as one
would expect for the same pressure jump in a 1D adiabatic shock.  This discrepancy
may be due to efficient radiative cooling or to additional wave effects caused by
shock bores. }
\end{center}
\end{figure}

\clearpage

\begin{figure}
\begin{center}
\includegraphics[width=5.95in]{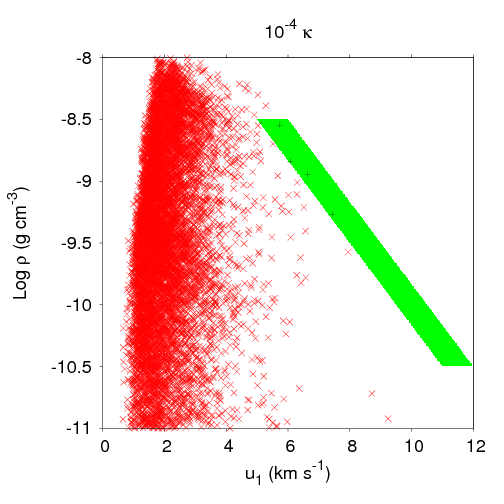}
\caption{Shocks on the $u_1$-$\rho$ plane for
the 1024 $10^{-4}~\kappa$ simulation, as determined from the fractional pressure change.  The green area
shows the chondrule-forming region, roughly based on the results of Desch \& Connolly
(2002).  The green strip should not be thought of as definitive, and slight changes
in the strip's location could identify more chondrule-forming shocks, particularly at low $u_1$.  If the
pre- and post-shock temperature ratio is used to determine $u_1$, no chondrule-forming shocks are detected and most shocks have a $u_1 < $ 4 km s$^{-1}$.}
\end{center}
\end{figure}


\begin{thebibliography}

\bibitem[Akima(1970)]{akima1970}
Akima, H.~1970, JACM, 17, 4

\bibitem[Armitage et al.(2001)]{armitage_etal_2001}
Armitage, P.~J., Livio, M., \& Pringle, J.~E.~2001, MNRAS, 324, 705

\bibitem[Balbus \& Hawley(1991)]{balbus_hawley1991}
{Balbus}, S.~A. \& {Hawley}, J.~F.~1991, ApJ, 376, 214

\bibitem[Bell \& Lin(1994)]{bell_lin_1994}
Bell, K.~R., \& Lin, D.~N.~C.~1994, ApJ, 427, 987

\bibitem[Binney \& Tremaine(1987)]{binney_tremaine}
Binney, J., \& Tremaine, S.~1987, Galactic dynamics (Princeton: Princeton Univ.~Press, 1987)


\bibitem[Bizzarro et al.(2004)]{bizzarro_etal_2004}
Bizzarro, M., Baker, J.~A., Haack, \& H.~2004, Nature, 431, 275

\bibitem[Boley(2007)]{boley_phd}
Boley, A.~C.~2007, Ph.D.~Thesis, Indiana University

\bibitem[Boley \& Durisen(2006)]{boley_durisen_2006}
Boley, A.~C., \& Durisen, R.~H.~2006, ApJ, 641, 534

\bibitem[Boley et al.(2005)]{boleyetal2005}
Boley, A.~C., Durisen, R.~H., \& Pickett, M.~K.~\chondrites, 839

\bibitem[Boley et al.(2007b)]{boleyetal2007b}
Boley, A.~C., Durisen, R.~H., Nordlund, \AA, \& Lord, J.~2007b, ApJ, 665, 1254

\bibitem[Boley et al.(2007a)]{boleyetal2007a}
Boley, A.~C., Hartquist, T.~W., Durisen, R.~H., \& Michael, S.~2007a, ApJ, 656, L89

\bibitem[Boss(1997)]{boss_1997}
Boss, A.~P.~1997, Science, 276, 1836

\bibitem[Boss(1998)]{boss_1998}
--.~1998, ApJ, 503, 923

\bibitem[Boss(2001)]{boss_2001}
--.~2001, ApJ, 562, 367

\bibitem[Boss(2002)]{boss_2002}
--.~2002, ApJ, 567, L149

\bibitem[Boss(2004a)]{boss_2004a}
--.~2004a, ApJ, 610, 456

\bibitem[Boss(2004b)]{boss_2004b}
--.~2004b, ApJ, 616, 1265

\bibitem[Boss(2005)]{boss_2005}
--.~2005, ApJ, 629, 535

\bibitem[Boss(2007)]{boss_2007}
--.~2007, ApJ, 661, L73

\bibitem[Boss(2008)]{boss_2008}
--.~2008, arXiv:0801.4371

\bibitem[Boss \& Durisen(2005)]{boss_durisen_2005a}
Boss, A.~P., \& Durisen, R.~H.~2005, ApJ, 621, L137

\bibitem[Cai et al.(2006)]{cai_etal_2006}
{{Cai}, K., {Durisen}, R.~H.,  {Michael}, S.,  {Boley}, A.~C., 
{Mej{\'{\i}}a}, A.~C., {Pickett}, M.~K., \& {D'Alessio}, P.}~2006, ApJ, 636, L149

\bibitem[Cai et al.(2008)]{cai_etal_2008}
Cai, K., Durisen, R.~H., Boley, A.~C., Pickett, M.~P., Mej\'ia, A.~C.~2007, ApJ, 673, 1138

\bibitem[Cameron(1978)]{cameron_1978}
Cameron, A.~G.~.W.~1978, M\&P, 18, 5

\bibitem[Ciesla \& Hood(2002)]{ciesla_hood_2002}
Cielsa, F.~J., \& Hood, L.~L.~2002, Icarus, 158, 281

\bibitem[Clarke et al.(2007)]{clarke_etal_2007}
Clarke, C.~J., Jarper-Clark, E., \& Lodato, G.~2007, MNRAS, 381, 1543

\bibitem[Cohl \& Tohline(1999)]{cohl_tohline_1999}
Cohl, H.~S., \& Tohline, J.~E.~1999, 527, 86

\bibitem[Cuzzi \& Alexander(2006)]{cuzzialexander2006}
Cuzzi, J.~N., \& Alexander, C.~M.~O'D.~2006, Nature, 441, 483

\bibitem[D'Alessio et al.(2001)]{dalessio_etal_2001}
{{D'Alessio}, P., {Calvet}, N., \& {Hartmann}, L.}~2001, ApJ, 553, 321

\bibitem[D'Alessio(2006)]{dalessio_etal_2006}
D'Alessio, P., Calvet, N., Hartmann, L., Franco-Hern\'andez, R., \& Serv\'in, H.~2006, ApJ, 638, 314

\bibitem[Desch(1998)]{desch_1998}
Desch, S.~J.~1998, Ph.D.~Thesis, University of Illinois at Urbana-Champaign

\bibitem[Desch(2004)]{desch_2004}
--.~2004, ApJ, 608, 509

\bibitem[Desch(2007)]{desch_2007}
--.~2007, ApJ, 671, 878

\bibitem[Desch \& Connolly(2002)]{deschconnolly2002}
Desch, S.~J., \& Connolly, H.~C., 
Jr.~2002, M\&PSA, 37, 183

\bibitem[Durisen et al.(2007)]{durisen_ppv}
Durisen, R.~H., Boss, A., Mayer, L., Nelson, A., Quinn, T., \& Rice, K.~2007,
in Protostars and Planets V, ed.~B.~Reipurth, D.~Jewitt, \& K.~Keil (Tucson: Univ. Arizona Press), 607

\bibitem[Durisen et al.(2008)]{durisen_etal2008}
Durisen, R.~H., Hartquist, T.~W., \& Pickett, M.~K.~2008, ApJ, in press

\bibitem[Durisen et al.(1986)]{durisen_1986}
Durisen, R.~H., Gingold, R.~A., Tohline, J.~E., \& Boss, A.~P.~1986, ApJ, 305, 281

\bibitem[Fleming \& Stone(2003)]{flemming_stone2003}
Fleming, T.~P., \& Stone, J.~M.~2003, ApJ, 585, 908

\bibitem[Gammie(1996)]{gammie_1996}
Gammie, C.~F.~1996, ApJ, 457, 355

\bibitem[Gammie(2001)]{gammie_2001}
--.~2001, ApJ, 553, 174

\bibitem[Green et al.(2006)]{green_etal_2006}
Green, J.~D., Hartmann, L., Calvet, N., Watson, D.~M., 
Ibrahimov, M., Furlan, E., Sargent, B., \& Forrest, W.~J.~2006, ApJ, 648, 1099

\bibitem[Harker \& Desch(2002)]{harkerdesch2002}
Harker, D.~E., \& Desch, S.~J.~2002, ApJ, 565, 109

\bibitem[Hartmann(2006)]{hartmann_etal_2006}
Hartmann, L., D'Alessio, P., Calvet, N., \& Muzerolle, J.~2006, ApJ, 648, 484

\bibitem[Hartmann \& Kenyon(1996)]{hartmann_kenyon_1996}
Hartmann, L., \& Kenyon, S.~J.~1996, ARA\&A, 34, 207

\bibitem[Hayashi(1985)]{hayashi_etal_1985}
Hayashi, C., Nakazawa, K., Nakagawa, Y.~1985, in Protostars and Planets II, 
ed.~D.~C.~Black, \& M.~S.~Matthews (Tucson: Univ. Arizona Press), 1100

\bibitem[Iida et al.(2001)]{iida_etal_2001}
Iida, A., Nakamoto, T., Susa, H., \& Nakagawa, Y.~2001, Icarus, 153, 430

\bibitem[Johnson \& Gammie(2003)]{johnson_gammie_2003}
Johnson, B.~M., \&~Gammie, C.~F.~2003, ApJ, 597, 131

\bibitem[Klahr \& Bodenheimer(2003)]{klahr_bodenheimer2003}
Klahr, H.~H., \& Bodenheimer, P.~2003, ApJ, 582, 869

\bibitem[Kley \& Lin(1999)]{kley_lin_1999}
Kley, W., \& Lin, D.~N.~C.~1999, ApJ, 518, 833

\bibitem[Krot et al.(2005)]{krot_etal_2005}
Krot, A.~N., Scott, E.~R.~D., \& Reipurth, B.~2005, Chondrites and the Protoplanetary Disk, 
ASPC Series, 341.

\bibitem[Krumholz et al.(2007)]{krumholz_etal_2007}
Krumholz, M.~R., Klein, R.~I., McKee, C.~F., \& Bolstad, J.~2007, ApJ, 667, 626

\bibitem[Krumholz et al.(2007)]{krumholz_etal_2007}
Krumholz, M.~R., Klein, R.~I., \& McKee, C.~F.~2007, ApJ, 656, 959

\bibitem[Kuiper(1951)]{kuiper1951}
Kuiper, G.~P.~1951, PNAS, 37, 1

\bibitem[Levison et al.(2007)]{levison_etal_2007}
Levison, H.~F., Morbidelli, A., Van Laerhoven, C., Gomes, R., \& Tsiganis, K.~2007, arXiv:0712.0553

\bibitem[Lynden-Bell \& Kalnajs(1972)]{Lynden-bell_kalnajs_1972}
Lynden-Bell, D., \& Kalnajs, A.~J.~1972, MNRAS, 157, 1

\bibitem[Mayer et al.(2007)]{mayer_etal_2007}
Mayer, L., Lufkin, G., Quinn, T., \& Wadsley, J.~2007, ApJ, 661, 77

\bibitem[McKeegan et al.(2006)]{mckeegan2006}
{McKeegan}, K.~D., et al.~2006, 
	Science, 314, 1724
	
\bibitem[Mej\'ia(2004)]{mejia_phd}
Mej\'ia, A.~C.~2004, Ph.D.~Thesis, Indiana University

\bibitem[Mej\'ia et al.(2005)]{mejia_etal_2005}
Mej\'ia, A.~C., Durisen, R.~H., Pickett, M.~K., \& Cai, K.~2005, ApJ, 619, 1098
	
\bibitem[Miura \& Nakamoto(2006)]{miuranakamoto2006}
Miura, H., \& Nakamoto, T.~2006, ApJ, 651, 1272

 \bibitem[Nelson(2006)]{nelson_2006}
Nelson, A.~F.~2006, MNRAS, 373, 1039

\bibitem[Nelson et al.(2000)]{nelson_etal_2000}
Nelson, A.~F., Benz, W., \& Ruzmaikina, T.~V.~2000, ApJ, 529, 357
	
\bibitem[Oishi et al.(2007)]{oishietal2007}
Oishi, J.~S., Mac Low, M., Menou, K.~2007, arXiv:astro-ph/0702549


\bibitem[Osorio et al.(2003)]{osorio_etal_2003}
Osorio, M., D'Alessio, P., Muzerolle, J., Calvet, N., \& Hartmann, L.~2003, ApJ, 586, 1148

\bibitem[Picket et al.(1998)]{pickett_etal_1998}
Pickett, B.~K., Cassen, P.~M., , Durisen, R.~H.,  \& Link, R.~P.~1998,
ApJ, 504, 468

\bibitem[Pickett et al.(2000)]{pickett_etal_2000}
--.~2000a, ApJ, 529, 1034

\bibitem[Pickett et al.(2003)]{pickett_etal_2003}
Pickett, B.~K., Mej\'ia, A.~C., Durisen, R.~H., Cassen, P.~M., Berry, D.~K., \& Link, R.~P.~2003, ApJ,
590, 1060

\bibitem[Pollack et al.(1994)]{pollack1994}
Pollack, J.~B., Hollenbach, D., Beckwith, S., Simonelli, D.~P.,
Roush, T., \& Fong, W.~1994, 421, 615

\bibitem[Press et al.(1986)]{press_etal1986}
Press, W.~H., Flannery, B.~P., \& Teukolsky, S.~A.~1986, Numerical recipes.
The art of scientific computing (Cambridge: University Press)

\bibitem[Rafikov(2005)]{rafikov_2005}
Rafikov, R.~R.~2005, ApJ, 621, L69

\bibitem[Rafikov(2007)]{rafikov_2007}
--.~2007, ApJ, 662, 642

\bibitem[Rice et al.(2003)]{rice_etal_2003}
Rice, W.~K.~M, Armitage, P.~J., Bate, M.~R., \& Bonnell, I.~A.~2003, MNRAS, 339, 1025

\bibitem[Rice et al.(2005)]{rice_etal_2005}
Rice, W.~K.~M, Lodato, G., \& Armitage, P.~J.~2005, MNRAS, 364, L56

\bibitem[Roberts et al.(1979)]{roberts_etal_1979}
Roberts Jr., W.~W., Huntely, J.~M., \& van Albada, G.~D.~1979, ApJ, 233, 67 

\bibitem[Russell et al.(2005)]{russell_etal_2005}
Russell, S.~S., Krot, A.~N., Huss, G.~R., Keil, K., 
Itoh, S., Yurimoto, H., \& Macpherson, G.~J.~\chondrites, 317
	
\bibitem[Sano et al.(2000)]{sano_etal_2000}
Sano, T., Miyama, S.~M., Umebayashi, T., \& Nakano, T.~2000, ApJ, 543, 486	

\bibitem[Stamatellos et al.(2007)]{stamatellos_etal_2007}
Stamatellos, D., Whitworth, A.~P., \& Ward-Thompson, D.~2007, MNRAS, 379, 1390

\bibitem[Stamatellos \& Whitworth(2008)]{stamatellos_whitworth2008}
Stamatellos, D., \& Whitworth, A.~P.~2008, A\&A, in press
	
\bibitem[Stepinski(1992)]{stepinski_1992}
Stepinski, T.~F.~1992, Icarus, 97, 130

\bibitem[Tomley et al.(1991)]{tomley1991}
Tomley, L., Cassen, P., \& Steiman-Cameron, T.~1991, ApJ, 382, 530

\bibitem[Tomley et al.(1994)]{tomley1994}
Tomley, L., Steiman-Cameron, T.~Y., Cassen, P.~1994, ApJ, 422, 850

\bibitem[Toomre(1964)]{toomre_1964}
Toomre, A.~1964, ApJ, 139, 1217
	
\bibitem[Wood(1963)]{wood_1963}
Wood, J.~A.~1963, Icarus, 2, 152 

\bibitem[Wood(1996)]{wood_1996}
--.~1996, Meteoritics Planet.~Sci., 31, 641

\bibitem[Wood(2005)]{wood_2005}
--.~\chondrites, 953

\bibitem[Wooden(2005)]{woodenetal2005}
Wooden, D., Harker, D.~E., \& Brearley, A.~J.~\chondrites, 774

\bibitem[Zhu et al.(2007)]{zhu_etal_2007}
Zhu, Z., Hartmann, L., Calvet, N., Hernandez, J., Muzerolle, J., 
\& Tannirkulam, A.~2007, ApJ, 669, 483

\end{thebibliography}
\end{document}